# The EDSGC – VII: The Edinburgh–Milano Cluster Redshift Survey *

C.A. Collins[1], L. Guzzo[2], R.C. Nichol[3], S.L. Lumsden[4]
[1] *Astrophysics Group, School of Chemical and Physical Sciences, Liverpool John Moores University, Byrom Street, Liverpool, L3 3AF U.K. – cac@star.livjm.ac.uk*
[2] *Osservatorio Astronomico di Brera, Via Bianchi 46, I-22055 Merate, Italy – guzzo@astmim.mi.astro.it*
[3] *Department of Astronomy and Astrophysics, University of Chicago, 5640 S. Ellis Rd., Chicago, Il 60637, USA – nichol@huron.uchicago.edu*
[4] *Anglo-Australian Observatory, Epping Laboratory, PO Box 296, Epping NSW 2121, Australia – sll@aaoepp2.aao.gov.au*

## ABSTRACT

In this paper we present the redshifts of the galaxies and galaxy clusters which form the Edinburgh–Milano (EM) cluster redshift survey. A total of 777 galaxy redshifts have been measured in 94 clusters extracted from the digitised Edinburgh Durham Cluster Catalogue. We also present the radial velocity dispersions for 37 clusters. Observational and data reduction techniques are discussed in detail, together with the strategy adopted to determine the mean redshift of a cluster and to identify and discard plausible phantom clusters. Some 10% of our clusters show heavy contamination, indicating that projection is a serious problem for optically selected rich clusters. The median velocity dispersion estimated for a sub-sample of richness $R \geq 1$ clusters is $742 \pm 63 \,\mathrm{km\,s^{-1}}$. From a simple comparison with $\Omega = 1$ Cold Dark Matter models of structure formation, these results favour a biasing parameter $b = 1.6 - 2.0$ and are inconsistent with a bias outside of the range $b = 1.3 - 2.5$.

## 1 INTRODUCTION

Clusters of galaxies represent one of the most important probes of the universe on cosmologically interesting scales. With typical separations of $\simeq 10 - 20\, h^{-1}\,\mathrm{Mpc}$, their spatial distribution can be used to map the underlying matter distribution on scales which can discriminate between the various models of structure formation (e.g. Mann, Peacock & Heavens 1994; Croft & Efstathiou 1994; Borgani et al. 1995). These models can also be constrained by examining the dynamical properties of clusters (e.g. Frenk et al. 1990). The Abell catalogue (Abell 1958) and its southern counterpart (Abell, Corwin & Olowin 1989) have been the most widely used source of rich galaxy clusters. As such, the primary cluster redshift surveys used to study large-scale structure in the universe have concentrated on Abell clusters. To date, the largest single compilation of Abell clusters is that of Struble & Rood (1991a), who list 758 redshifts and 121 velocity dispersions, although this compilation ignores southern ACO and supplementary clusters.

One of the major scientific investigations using such samples of Abell clusters has been the determination of the amplitude of the clustering pattern of rich clusters. Bahcall & Soneira (1983) studied the clustering of a statistical sample of 104 clusters which constitutes all the high galactic latitude $R \geq 1$ and $D \leq 4$ Abell clusters, by estimating the two-point cluster correlation function $\xi_{cc}(r)$. More recently, Postman, Huchra & Geller (1992) have estimated $\xi_{cc}(r)$ for a complete sample of 351 Abell clusters with tenth ranked galaxy magnitudes $(m_{10}) \leq 16.5$. The results from these two studies are consistent in showing a correlation function of the form $\xi_{cc}(r) = (r/r_0)^{-1.8}$, with a correlation length $r_0 \geq 20\, h^{-1}\,\mathrm{Mpc}$. In addition, Peacock & West (1992) have carried out a power-spectrum analysis of a volume limited all-sky cluster sample containing 427 Abell clusters, which is $\sim 90\%$ complete in redshift for richness class $R \geq 0$. This study implies a similar correlation length.

However, these results have been challenged due to alleged systematic errors in the Abell catalogue, which could artificially enhance the value of $r_0$ (Lucey 1983, Sutherland 1988, Dekel et al. 1989, Erfstathiou et al. 1992). These systematics could be due to selecting clusters visually from photographic plates, or patchy Galactic extinction, or contamination of galaxy associations — which are simply projected enhancements along the line-of-sight. A significant step towards a more objective understanding of the nature of the galaxy distribution has been realised with the construction of digitised galaxy catalogues, such as the Edinburgh/Durham Southern Galaxy Catalogue (EDSGC, Heydon–Dumbleton et al. 1989) and the APM catalogue (Maddox, Efstathiou & Sutherland 1990). From these surveys, it has been possible to select clusters automatically and produce cluster samples which are not prone to the subjective effects discussed above. Specifically, from the EDSGC we have constructed the Edinburgh/Durham Cluster Catalogue (EDCC), consisting of 737 clusters over an area of 1500 deg$^2$, centered at the South Galactic Cap (Lumsden et al. 1992). The EDCC is complete to b$_j$ = 18.75, approximately 1 magnitude fainter than the Abell catalogue. A comparison with the Abell cluster catalogue in the same region of sky reveals that to the completeness limit of the EDCC, 80% of Abell clusters are detected. All aspects of the catalogue construction and a detailed listing of the clusters themselves are given in Lumsden et al. (1992).

---

\* Based mainly on data collected at the European Southern Observatory, La Silla, Chile.

During the last five years we have been measuring redshifts for galaxies in the EDCC. Our aim was to assemble a complete richness–limited spatial sample of clusters, with accurately known redshifts. To this end, a mean number of $\sim$ 10 galaxies per cluster was observed, a strategy that characterizes our survey and allowed a further check of residual projections. This survey is called the Edinburgh/Milano (EM) cluster redshift survey, and it is presented in detail in this paper. Previous papers related to the analysis of this data set have discussed the large–scale distribution (Guzzo et al. 1992), the correlation function of the clusters (Nichol et al. 1992), and large–scale cluster alignments (Martin et al. 1995). These analyses have indicated a lower level of clustering of rich clusters than implied by the Abell catalogue, with a correlation length $r_0 \simeq 15\,h^{-1}\,$Mpc. In addition to the data, we present here also an analysis of the velocity dispersion for a subset of 37 clusters.

The paper is organized as follows. In section 2 we describe in detail the observational procedures adopted at ESO and the AAT. In section 3 we discuss the reduction of the data to 1–D spectra and in section 4 we explain the galaxy redshift determinations. In section 5 we describe the determination of each cluster redshift, along with an explanation of how we identify clusters as projection effects. In section 6 we present the cluster velocity dispersions and briefly discuss their implications for cosmology. In section 7 the EDCC cluster sample used to determine the correlation function is defined, as the exact selection criteria used for the correlation function sample differs from the manner in which the EDCC was selected in certain important respects.

## 2 OBSERVATIONS

### 2.1 Observing Strategy

In adopting a strategy for taking the redshift of clusters, our main concern was the problem of galaxy interlopers and phantom clusters caused by the alignment, along the line-of-sight, of small groups of galaxies. Lucey (1983) argues that the Abell catalogue contains a significant fraction of heavily contaminated clusters. His models indicate that between 15 − 25% of the clusters in the Abell catalogue have a true galaxy population that is less than half of their observed population. Although there are counter arguments suggesting that the contamination is not that significant (Struble & Rood 1991b), this point can only really be addressed by taking multiple redshift measurement towards the cores of the clusters. The majority of existing cluster redshift surveys are based mainly on one or two galaxy redshift measurements per cluster. For example, $\sim$ 68% of the cluster redshifts in the Struble & Rood (1991a) compilation are based on < 3 galaxies and the two largest systematic cluster redshift surveys to date, the Abell-based survey of Postman et al. (1992) and the APM survey (Dalton et al. 1994), contain significant fractions of single or double cluster redshift measurements. The APM survey contains 188 clusters, of which $\sim$ 70% are based on 1 or 2 galaxy redshifts; while for the Postman sample, the figure is $\sim$ 50%. Phantom clusters cannot be detected with so few redshifts, while for real clusters, there is a finite chance that the redshift of an interloper will be measured and assigned to the cluster. Therefore, the strategy behind the construction of the EM redshift survey was to measure about 10 galaxy redshifts per cluster towards the centres of the rich clusters selected from the EDCC.

### 2.2 ESO Observations

Over 70% of all the observations made in the construction of the EM survey were carried out using the ESO Faint Object Spectrograph and Camera (EFOSC) on the 3.6 m telescope at La Silla, Chile. EFOSC is a high-efficiency transmission spectrograph, with multi-object spectroscopic capability (MOS) and fast switching to imaging mode. This latter feature allows very accurate slit positioning on faint objects. At the time of our observations (1988–1991), the detector was a thinned, back–illuminated RCA CCD chip of size 520 × 320 pixels, with good response in the blue band. The pixel size with this setup is 30 $\mu$m, corresponding to 0.675 arcsec on the sky and the dispersion direction runs along the longest CCD axis. The chip is cosmetically clean with a read–out noise of 45 electrons. The peak quantum efficiency of the chip is reached at 4800 Å and remains > 60% over the observed wavelength range.

EFOSC was mainly used in MOS mode, which entailed producing masks of the clusters which were then inserted into free positions in the aperture wheel of the spectrograph (see below). For single slit observations, we used a 2–arcsec wide slit, the same width obtainable on the MOS masks. We used the B300 grism which has a dispersion of 230 Å/mm. With the RCA chip this corresponds to a resolution of 6.9 Å per pixel, providing a wavelength coverage between 3600 Å − 7000 Å. With this setup, spectral resolution as measured on a purely instrumentally broadened line resulted in about 2 pixels FWHM. Using a cross–correlation technique, this moderate resolution enabled us to achieve an rms external error on the radial velocities of $\sim$ 140 km s$^{-1}$ (see section 4.2).

MOS operations with EFOSC involve the use of pre-prepared masks on which slitlets of length 5 – 30 arcsec are punched. Their positions on the mask are determined from a direct CCD image of the cluster. On average each mask contained 10 – 15 slitlets, with a maximum of 22 for the cluster E429. The integrations were carried out in two consecutive exposures each of 20 – 30 minutes duration which facilitated the removal of cosmic–ray events. He–Ar calibration spectra were observed before and after the science exposures through the same mask. A few clusters, not suitable for MOS were observed in single–slit mode. Standard bias frames and flat–fields were taken at the beginning and end of each night.

A limited number of redshifts were obtained at the ESO 2.2 m telescope. For these observations a classical Boller & Chivens spectrograph coupled with an RCA CCD was used, with a spectral setup similar to that adopted for EFOSC. During these runs we also re–observed some of the clusters already secured with the 3.6 m, in an effort to establish external redshift errors (see section 4.2).

## 2.3 AAT OBSERVATIONS

About 30% of the survey data were collected during a 3–night run at the Anglo–Australian 3.9 m Telescope (AAT) using the multi–fibre spectrograph AUTOFIB coupled to the RGO spectrograph. This instrument has 60 user fibres that are automatically positioned by a flying robot over a circular field of 40 arcmin in diameter at the Cassegrain focus of the AAT. Full instrumental details are given in Parry & Sharples (1988). The optimum instrumental combination proved to be the 600V grating in conjunction with the IPCS. However, due to technical constraints, half the clusters were observed using the GEC CCD. Both detectors provide a wavelength coverage 3700 Å – 5500 Å, with a resolution of 3 Å per pixel for the CCD and 1.7 Å for the IPCS.

The positions of the target galaxies, to be fed into the system controlling AUTOFIB for positioning the 2 arcsec fibres, were taken from the EDSGC, which provides a positional accuracy of $\simeq$ 1 arcsec. The galaxy target list was constructed by giving priority to bright galaxies near the centre of the clusters. Within the mechanical constraints we observed 30 – 40 galaxies per cluster, with the remaining fibres placed on the sky near concentrations of object fibres. The sequence of object and calibration exposures was the same as that adopted for the ESO observations. Finally, for both the ESO and AAT observations, radial velocity standard stars were observed on most nights.

## 3 DATA REDUCTION

### 3.1 EFOSC Data

After subtracting the bias, the two science exposures available for every MOS observation were averaged together using a k–sigma clipping algorithm, which very effectively removed most of the cosmic ray events. The data were reduced using the MIDAS package. When necessary, appropriate programmes were developed to facilitate the reduction of the multi–object data (extraction of single spectra, wavelength calibration and sky subtraction). The lamp flat–fields were normalized by removing the large–scale spectral response of the lamp before being used, leaving only the pixel–to–pixel variations. These proved to be very small ($\simeq$ 2%) and had negligible effect on the estimate of line positions. For each MOS frame, the first arc spectrum was searched for a number of calibration lines, a few of which were manually identified. Supplemented by an automatic search, this typically resulted in $\simeq$ 15 lines which were used to calculate the pixel–to–wavelength relationship to be applied to the corresponding science spectrum. The typical $rms$ wavelength calibration error using a 3rd order polynomial was $\sim$ 0.3 Å. The object spectrum was then rebinned into constant wavelength steps of 5 Å. A further test of the accuracy of the wavelength calibration was performed by checking the wavelength of sky emission lines from the calibrated spectrum against their nominal value.

Once the spectrum from the first slitlet was calibrated, the approximate zero–point shift of the following arc in the same MOS exposure was found by pointing to a bright Helium line. This was used by the program as a first guess to re–identify the calibration lines and find a new wavelength solution for the second spectrum. The process continued until all the spectra in the MOS frame were calibrated. The whole procedure could be checked in real time and, if necessary, paused to correct for misidentified lines. The resulting calibrated spectra were then reduced to one dimension and sky subtracted. Since some spectra did not have sky on the slit, performing this operation after wavelength calibration allowed us to use the sky from an adjacent slit when necessary. For most of the objects sky subtraction was very effective (a clear advantage of slitlets compared to fibres). All the resulting 1–D spectra were directly inspected and sky-line residuals removed by linear interpolation of the galaxy continuum across the line.

Over the six ESO observing runs, a total of $\simeq$ 544 galaxy spectra were observed. Fig. 1 shows some typical galaxy spectra selected from the ESO database.

### 3.2 AUTOFIB Data

It was decided not to flat–field either the CCD or IPCS data, as the pixel to pixel response of both detectors had negligible effect on the line positions. Fibre spectra were then extracted from the parent frames using dedicated commands within the FIGARO package. This automatically applied a correction for S distortion in the dispersion direction. The final product was a reduced 2–D data frame, within which each of the 55 – 60 rows corresponded to a different spectrum. The arcs taken for these data proved to contain only low signal–to–noise lines. In order to increase the signal to noise, we explored the possibility of co–adding all the calibration arcs taken on the same night. It was found that the $rms$ temporal shifts in the arcs was $\sim$ 0.05 pixels, which is one order of magnitude lower than the expected $rms$ calibration residuals. Given this negligible shift, we constructed a master arc for each night to calibrate all the spectra in each 2–D frame. The master arc provided 40 – 50 identified lines spread evenly throughout the wavelength window. The calibration relation was then obtained by a 4th order polynomial fit, which gave a typical $rms$ residual of $\sim$ 0.3 Å for the IPCS data and $\sim$ 0.7 Å for the CCD data. As a final verification of the accuracy of the wavelength calibration, we checked the stability of the [O I] $\lambda 5577$ Å sky line over the whole night. This was found to be at the correct wavelength with an *rms* deviation similar to that quoted above.

There is no generally agreed optimal procedure for sky-subtraction using fibres (see Ellis & Parry 1988). One possibility is to dedicate some fibres to the sky only and then renormalize their relative transmission. This can be achieved by measuring 'offset skies' before or after the science exposure by offsetting the telescope by a small amount and taking a short sky exposure through the configured fibres. With this method the degree of bending and torsion in the fibres remains close to that in the science exposure thereby preserving the relative transmission. However, to obtain a sufficient signal-to-noise ratio, sky exposures $\geq 15$ minutes duration are required and for this reason we did not adopt this strategy. Instead, we constructed a median sky spectrum from all the sky fibres, after dividing each single sky fibre spectrum by its median count. Then we used the ratio of the heights of the [O I] $\lambda 5577$ Å sky lines to scale the median sky to each fibre response and subtracted. This technique proved to be very successful (see Nichol 1992), and has been recently discussed in detail in other papers (e.g. Lissandrini, Cristiani & La Franca 1994; Ettori, Guzzo & Tarenghi 1995).

## 4 GALAXY REDSHIFT DETERMINATION

To estimate the redshifts from the calibrated spectra we used the cross-correlation technique described in detail by Tonry & Davis (1979), in the version developed within FIGARO which closely follows the original prescription. The basis of the technique is the cross-correlation of the observed galaxy spectrum with a model or template spectrum. This is performed by taking the Fast Fourier Transform of the two spectra, multiplying them together and then transforming back the result to get the Cross-Correlation Function (CCF), whose highest peak is related to the radial velocity difference between the two spectra. Before actually starting this machinery, the two spectra are rebinned into logarithmic bins, so that the relative redshift becomes a linear shift between the two and a number of operations are performed on them to improve the quality of the final result. These are described in detail by Tonry & Davis (1979), and include continuum subtraction, cosine-bell apodizing and bandpass filtering. All these operations were performed using FIGARO specific commands, and several combinations of the parameters involved were tested to find the most appropriate set for our spectra. In particular, the spectra were rebinned into 2048 logarithmic bins. For the ESO data, this corresponded to a velocity binwidth of $84.5\,{\rm km\,s^{-1}}$, while for the AUTOFIB data it was $57.9\,{\rm km\,s^{-1}}$. The redshift of the galaxy was found to be insensitive to the exact binwidth chosen (2048 or 4096). The spectra were then filtered in order to eliminate both the low frequency spurious components left by the subtracted continuum, and the high frequency binning noise. The best set of filter parameters was chosen to maximize the significance of the CCF. The peak of the CCF was fit by a quadratic polynomial, determining both the wavelength shift, from its position, and the error, from its width.

### 4.1 Template Spectra

During the course of the project, 27 templates were observed, 17 at ESO and 10 at the AAT. Fifteen of the ESO templates were galaxies, some of which were late-type galaxies with very accurate 21cm line redshifts (errors $\sim 10\,{\rm km\,s^{-1}}$, da Costa *et al.* 1984). These were used to set the zero point for the whole set of templates. The other 2 ESO templates were radial velocity standard stars of spectral type K, taken from the Astronomical Almanac. The AUTOFIB templates were all standard stars from the Astronomical Almanac, however these were discarded as the ESO templates were of a higher quality. To check the zero-point reliability of the templates, they were cross-correlated against each other (including the 21cm galaxies), to obtain the redshift of each template with respect to the other 16. For each object, the difference between its published redshift and the measured redshifts from the other 16 templates was calculated. Overall, seven of the original seventeen ESO templates were classified as having very reliable zero-points, and selected to be used in the cross-correlation technique. Two more archive templates were chosen on the basis of their previous reliability and high signal-to-noise (Parker, Beard, & MacGillivray 1987). Finally, we chose as a further template, the high signal-to-noise spectrum of a galaxy in A4038. This was meant to provide at least one potentially good AAT template. In reality, the ESO templates proved to be superior, with no visible systematic effect. The ten final templates are listed in Table 1 along with their published redshifts.

### 4.2 Cross-Correlation and Errors

The major advantage of the cross-correlation technique is that it preserves the statistical information on the redshift contained in the whole spectrum. A useful feature of the cross-correlation package used was the ability to place a confidence level on the chosen CCF peak by comparing its height with that expected for a random noise peak in the CCF. This followed the Tonry & Davis prescription (see Heavens 1993 for a new, more accurate approach to the problem). We introduced into the cross-correlation program a confidence level estimate on the redshifts obtained, based on the stability of the result provided by the different templates. Each galaxy spectrum was cross-correlated against the 10 templates described above and a confidence level was assigned to each cross-correlation. Spectra with 5 or more templates in agreement (within the errors on the redshifts) with confidence levels above 0.95 ($\simeq 2\sigma$) were passed as secure. Spectra that did not satisfy these criteria were inspected by eye. Most of these spectra had between 2 and 4 templates in agreement and the visual inspection often supported this lower level of agreement. Spectra were discarded if there was no agreement between the templates and a visual inspection could not determine the redshift. Once a spectrum had been accepted as secure (visually or with $\geq 5$ templates), the template redshift with the highest confidence level was assigned to the galaxy. If several templates had the same confidence, then the one with the lowest returned internal error was used. For high signal-to-noise spectra, it was common to find all the templates agreeing to within with a scatter of $\Delta v \simeq 50\,{\rm km\,s^{-1}}$. Lower signal-to-noise spectra had a typical scatter of $\Delta v \simeq 100\,\mathrm{km\,s^{-1}}$. A small fraction of the galaxies had emission lines in their spectra. However, in most cases the emission lines did not provide redshifts of sufficient accuracy ($\delta v \simeq 250\,\mathrm{km\,s^{-1}}$) and these were only used when no absorption redshift was available.

Following the prescription of Tonry & Davis (1979), the cross–correlation programme provides an estimate of the internal error on the redshift, based on the width of the cross–correlation peak. For our data this ranged between $14\,\mathrm{km\,s^{-1}}$ and $683\,\mathrm{km\,s^{-1}}$.

Throughout the course of the project, a total of 30 galaxy redshifts were repeated. These repeat observations can be used to provide an estimate of the external error on the redshifts. Table 2 shows the redshifts for all the repeat measurements in the survey along with the telescope on which each redshift was measured. The median offset between all these repeats is $\simeq 200\,\mathrm{km\,s^{-1}}$, implying an $rms$ external error of $\simeq 140\,\mathrm{km\,s^{-1}}$. This value is representative of the error on all three telescopes on which the redshifts were measured. In Table 3 we present all the galaxy redshifts obtained in each cluster field as part of the EM survey.

## 5 CLUSTER REDSHIFTS

### 5.1 Determining Cluster Membership

The redshift of a cluster was defined in an objective manner which removed any subjective "eyeball" determination of whether a galaxy was in a cluster or not and provided a set of well–defined selection criteria that could be accurately reproduced. In addition, these criteria were used to quantify the contamination due to galaxy interlopers and used to determine the frequency of phantom clusters. We first experimented with the simple $3\sigma$ pessimistic clipping scheme adopted by Colless & Hewett (1987). This method worked well for most cases to remove single outliers but broke down in cases where the clusters had a significant fraction of galaxies away from the main concentration (e.g. E519). We therefore supplemented the clipping algorithm with criteria similar to those described by Struble & Rood (1991b), which can be outlined as follows:

(i) The mean and standard deviation of all the galaxy redshift measurements for a single cluster were calculated. The most discrepant galaxy redshift was temporarily removed from the galaxy listing and the mean redshift and standard deviation ($\sigma_r$) were re–calculated. If $\sigma_r$ was found to be outside the range $\sigma_f < \sigma_r < 4\sigma_f$, where $\sigma_f$ ($= 700\,\mathrm{km\,s^{-1}}$) is a fiducial median cluster velocity dispersion, as quoted by Zabludoff, Huchra & Geller (1990), then it was set to the nearest of these limits to take account of clusters with unrealistically large or small values of $\sigma_r$.

(ii) All the galaxy redshifts taken for a particular cluster were binned in redshift with a binwidth equal to $\sigma_f$.

(iii) The cluster redshift distribution was searched for its highest peak. Once found, other peaks in its vicinity were located and merged if they were $< \sigma_r$ away from the original peak. Various multiples of $\sigma_r$ were tried, but it was found empirically that a threshold of $\sigma_r$ gave the most realistic results. All the galaxy redshift measurements that had been merged into this one peak were written out and removed from the galaxy listing. This procedure was repeated until all the redshift peaks had been located, which produced a list of redshift concentrations for the cluster which could vary from a single galaxy to all the galaxies observed in that cluster.

(iv) For each concentration, the mean redshift and the fraction of observed galaxies within that concentration were calculated. The concentration with the highest fraction was taken as representative of the true cluster redshift.

(v) If a cluster had a secondary concentration within its distribution that contained more than a third of the observed galaxies and was separated by more than $1500\,\mathrm{km\,s^{-1}}$ $i.e. \simeq 2$ binwidths, from the largest concentration mentioned above, then the cluster was defined as a projection effect or phantom cluster. If the peaks were separated by less than $1500\,\mathrm{km\,s^{-1}}$, then they were merged together and re–analysed (Stage 4). This guarded against clusters with subclustering and/or high velocity dispersions being broken up and classed as spurious. All the remaining redshift concentrations for a cluster were defined as interlopers. These values of acceptance were found empirically.

If the number of residual galaxies left in a cluster after $3\sigma$ clipping was more than 2 larger than the results of the de–projection, then the latter result was adopted. For clusters with fewer than 5 members, only obvious outliers were removed, as the statistics on $\sigma_r$ were then too poor. The above algorithm is rather crude and in all probability could be improved. For example, by weighting each galaxy redshift by the inverse of the distance from the cluster centroid or using the distribution of apparent magnitudes for the galaxies (see Dalton et al. 1994). A notable example where our criteria may be too rigid is E127 (Abell 3856) which we class as a projection effect (fig. 2). It was observed at the AAT and thus has measured galaxies distributed over the 40 arcmin AUTOFIB field–of–view. The two galaxies within the cluster core plus another two more external ones share a similar redshift of $\sim 40000\,\mathrm{km\,s^{-1}}$, thus suggesting that value as the possible cluster redshift. The mean cluster velocity and standard error for all EM clusters are shown in Table 4.

### 5.2 Interlopers and Phantom Clusters

The objective criteria described above lead to a natural definition of an interloper and a phantom cluster and provide direct estimates of the contaminating percentages in the EDCC. To do this we confine ourselves to the ESO data only, since the smaller field–of–view constrains the observations to the cluster cores. For all the observed ESO EM clusters, the percentage of clusters that had any amount of interloper contamination is 65%. For these clusters, but excluding phantom ones, the percentage of sampled redshifts per cluster defined as interlopers is 27%. The number of spurious clusters (30% of members more than $1500\,\mathrm{km\,s^{-1}}$ away) out of the total 83 EM clusters observed at ESO is 8, $i.e.$ 10%. For the small–Abell–radius sample used in Nichol et al. (1992), 6 clusters were classified as phantom $i.e.$ 6% of that sample. These figures indicate that a single redshift taken towards the core of a cluster has a significant chance of not being a member of that cluster. In addition, $\sim 10\%$ of a richness which is overestimated by $\sim 30\%$.

It is interesting to compare these figures with those obtained for the Abell catalogue which rely on detailed modelling techniques. Our results are in reasonable agreement with Lucey (1983), who uses Monte Carlo simulations to estimate that $15\% - 25\%$ of Abell clusters are significantly contaminated ($\geq 50\%$). The value of $\simeq 50\%$ estimated from an analytical models by Fesenko (1979) is significantly higher than our estimates. Struble & Rood (1991b) used redshift measurements to estimate that only $3\% - 5\%$ of Abell clusters were superpositions and thus spurious. Our data would suggest that this is an under-estimate. Their low value might be due to the fact that they were mainly concerned with estimating the effects of interloper contamination on the richness of the cluster and whether a cluster could be boosted up into Abell's statistical sample. Therefore, they set out with the apriori assumption that all the "clusters" are clusters of some form or another, so their result is probably a lower limit.

### 5.3 Cluster Redshift Comparisons

Our data can be compared with several recent compilations of cluster redshifts. A total of 13 clusters are common with the recently published APM list of cluster redshifts based on 2 galaxies per cluster and a likelihood estimator (Dalton et al. 1994). Of these, 12 have an absolute average median velocity residual of $295\,\mathrm{km\,s^{-1}}$ (E261, E400, E410, E447, E462, E557, E683, E699, E735, E742, E748, E765). For one cluster the difference in methodology between many redshifts per cluster and a likelihood estimator to determine cluster redshifts becomes more apparent. For the cluster, E470, our redshift, based on 7 galaxies, is $32903\,\mathrm{km\,s^{-1}}$. Dalton et al. adopt a cluster redshift from their single measured galaxy redshift of $22784\,\mathrm{km\,s^{-1}}$, rather than their likelihood value of $32647\,\mathrm{km\,s^{-1}}$, which is in fact much closer to our value. There are 3 clusters in common with the clusters observed by Colless & Hewett (1987). The clusters E124, E394 and E400 have redshift differences (EDCC–C&H) of $-547\,\mathrm{km\,s^{-1}}$, $-45\,\mathrm{km\,s^{-1}}$ and $-334\,\mathrm{km\,s^{-1}}$ respectively. E400 has also been observed extensively by Teague, Carter & Gray (1990) who have measured the redshifts of 72 cluster members. There is a difference in these redshift estimates (EDCC–T&G) of $145\,\mathrm{km\,s^{-1}}$. The recent compilation of Abell redshifts published by Lauer & Postman (1994) has 4 clusters in common with our own. The clusters E348 and E394 have velocity differences (EDCC–L&P) of $258\,\mathrm{km\,s^{-1}}$ and $-28\,\mathrm{km\,s^{-1}}$ respectively. For the cluster E372 we have only one redshift of $12675\,\mathrm{km\,s^{-1}}$ for a galaxy located 4 arcmins from the Abell cluster centre. Lauer & Postman (1994) quote a redshift of $14739\,\mathrm{km\,s^{-1}}$ for this cluster. There is a larger discrepancy between the measured redshifts of E557. Lauer & Postman (1994) quote a redshift for this cluster (A2911) of $6074\,\mathrm{km\,s^{-1}}$. We have taken redshifts for 8 galaxies towards this cluster and 6 have a mean of $23729\,\mathrm{km\,s^{-1}}$ with a small velocity dispersion of $413\,\mathrm{km\,s^{-1}}$. This result is almost midway between the Dalton et al. (1994) measured redshifts of $23174\,\mathrm{km\,s^{-1}}$ and $24193\,\mathrm{km\,s^{-1}}$. Fig. 3 shows a right ascension cone diagram of all the EM clusters classified as genuine, including the literature clusters, providing a view of how these systems delineate the large-scale structure of the universe. The details of this distribution, and particularly the comparison with previous results from galaxy redshift surveys, have been discussed elsewhere (Guzzo et al. 1992, Nichol et al. 1992).

## 6 VELOCITY DISPERSIONS

### 6.1 Estimator and Results

The line-of-sight velocity dispersions have been estimated for all clusters in the sample which have 6 or more redshifts belonging to the cluster, on the above definition. The velocity dispersions have been calculated using the rigorous procedure described by Danese, De Zotti & di Tullio (1980). The radial velocity dispersion $\sigma_v$ is given by

$$\sigma_v = \sum_{i=1}^{N} v_i^2/(n-1) - \delta^2/(1+\overline{V})^2, \qquad (1)$$

where $\overline{V}$ is the mean cluster redshift ($\overline{V} \simeq c\overline{z}$), $\delta$ is the contribution from the measurement error, and $v$ is line-of-sight component of the velocity of a galaxy with respect to the cluster centre, given by

$$v = (V - \overline{V})/(1 + \overline{V}/c), \qquad (2)$$

where $V \simeq cz$ is the redshift of each galaxy. We have also calculated the statistical uncertainty in the velocity dispersions using the prescription given in Danese et al. (1980). This accurately takes into account sampling errors, experimental errors and cosmological corrections. The values of $\sigma_v$ and their errors for 37 EDCC clusters with a net number of galaxies per cluster $N_{clus} \geq 6$ are presented in Table 4.

There are three of our clusters in common with the Colless & Hewett (1987) sample of cluster velocity dispersions for which they have $N_{clus} \simeq 40$. The velocity dispersion differences (EDCC–C&H) for the 3 clusters E124, E394, E400 are $-169\,\mathrm{km\,s^{-1}}$, $51\,\mathrm{km\,s^{-1}}$ and $211\,\mathrm{km\,s^{-1}}$ respectively. Teague et al. (1990) find $\sigma_v = 885\,\mathrm{km\,s^{-1}}$ for E400 using $N_{clus} \simeq 70$ galaxies. This is $\simeq 1.5\,\sigma$ smaller than our estimate of $1222\,\mathrm{km\,s^{-1}}$ based on 7 galaxies.

### 6.2 Comparison with Cosmological Models

The velocity dispersions of clusters can be used to place conservative constraints on cosmological models. Frenk et al. (1990) (hereafter FWED) calculate the distribution of velocity dispersions for $R \geq 1$ clusters in several $\Omega = 1$ cold dark matter (CDM) cosmologies with different values for the biasing parameter b. In comparing our data with FWED, we follow closely the procedure of Zabludoff et al. (1990) and compare two statistics; the largest expected velocity dispersion in our sample and the median velocity dispersion of the whole sample. FWED present the velocity distributions for clusters selected from simulations carried out in 3–D (their Fig. 1). In addition, they argue that projection can seriously contaminate cluster selection leading to an over-estimate of the velocity dispersions. Therefore, they also model cluster selection in 2–D by subjecting cluster catalogues from the simulations to projection effects expected in the Abell catalogue and using techniques employed in the analysis of real clusters (their Fig. 3).

A histogram of the velocity dispersions for our sample of 37 clusters is shown in Fig. 4. The largest velocity dispersion in our sample of 36 is E400 (A2721) with $\sigma_v = 1222^{+606}_{-243}$ km s$^{-1}$. However, the comparison with the more accurate measurement of Teague, Carter & Gray (1990) reported in section 6.1, indicates this to be overestimated by $\sim 350$ km s$^{-1}$. The next highest is E606 (Abell 2962), with $\sigma_v = 1152^{+377}_{-190}$ km s$^{-1}$. Interestingly, these values are not a good fit to the Frenk et al. 2-D velocity dispersion model, which predicts too many clusters with high velocity dispersions ($\geq 25\%$ with $\sigma_v \geq 1200$ km s$^{-1}$ for b $\leq 2.5$). Comparing with the 3-D selection predictions, the only compatible CDM models are those with b $\geq 1.6$, which predict $\leq 10\%$ of clusters will have $\sigma_v \geq 1200$ km s$^{-1}$. For b = 1.3 the predicted fraction rises to 40% and can be ruled out.

In order to compare the median velocities of our sample with the predictions of CDM for R $\geq 1$ clusters we construct a sub-sample of 24 clusters which satisfy the criteria R $\geq 1$. For these clusters we derive a median cluster velocity dispersion of $742 \pm 63$ km s$^{-1}$. This agrees to within $1\sigma$ with the predictions of the median velocities for CDM with b = 1.6 – 2.0, viz. 850 km s$^{-1}$ and 760 km s$^{-1}$ respectively (Zabludoff et al. 1990). For b $\geq 2.5$ and b $\geq 3.3$, our value is too large by $\geq 2\sigma$ and $\geq 4\sigma$ respectively. If b $\leq 1.3$, then the predicted median velocity rises to $\geq 1000$ km s$^{-1}$ and is firmly inconsistent with the data.

In two independent comparisons the same conclusion can be reached; namely, the EM velocity dispersion data are most consistent with a b = 1.6 or a b = 2.0 CDM model and can strongly exclude models with b $\geq 2.5$ and b $\leq 1.3$. These conclusions are in substantial agreement with those of Zabludoff et al. (1990) and Girardi et al. (1993), both of whom compare the distribution function of velocity dispersions for Abell cluster samples with the predictions of FWED. It should be taken into account, however, the possibility of a physical velocity bias which can lower the observed velocity dispersion in clusters to about 70% that of the dark matter. As shown by Couchman & Carlberg (1992), this would make a low bias (b=1 or less) CDM model, compatible with the COBE-DMR measurements, to agree with basically all the current observational data, including those presented here.

## 7 THE CLUSTER–CLUSTER CORRELATION FUNCTION SAMPLE

One of the main motivations behind the construction of the EM survey was to re-estimate the cluster spatial autocorrelation function $\xi_{cc}(r)$ from a sample where systematic errors and biases that had allegedly plagued previous estimates could be kept under control. As mentioned in Section 1, the major concern regarding the Abell catalogue was the extent to which it is seriously contaminated by projection effects. One of the easiest ways of reducing projection effects is to reduce the size of the counting radius within which the cluster is defined. This reduces the number of cluster overlaps and thus prevents the richness of distant clusters, which are in the haloes of nearby clusters, being overestimated and consequently being incorrectly included into the catalogue. In the construction of the Abell catalogue, a radius of $1.5\,h^{-1}$ Mpc was used to define the clusters. Many authors believe this is an overestimate for the size of clusters and even George Abell, in his original paper (Abell 1958), commented that his radius may be too large. The EDCC was constructed using the standard Abell radius mentioned above, as this allowed for a fair comparison between it and the Abell catalogue (Lumsden et al. 1992). However, for the above reasons, for the computation of the cluster correlation function a new sample of clusters was selected from the EDCC using a smaller radius of $1.0\,h^{-1}$ Mpc. This reduced the number of deblended clusters from 30% in the standard EDCC to 8% in this sample (Nichol 1992). The final redshift sample used to calculate the correlation function discussed in Nichol et al. (1992) was thus selected using the following 3 criteria:

(i) A background corrected $m_{10}(b_j)$ of $\leq$ 18.75, which corresponds to the completeness limit of the EDCC (z$\sim 0.13$).
(ii) Within the coordinate limits of $21^h.88 <$ RA $< 3^h.59$ and $-42°.40 <$ Dec $< -22°.88$. This prevented clusters near the edge of the survey being included as their statistics were often uncertain.
(iii) Clusters with a galaxy richness, $R_1 \geq 22$ galaxy members within an Abell radius of $1.0\,h^{-1}$ Mpc, after background correction.

¿From a comparison of the clusters in common between this sample and the Abell catalogue, the $R_1 \geq 22$ richness cut corresponds to a richness cut of 40 for Abell clusters. This means that this small-Abell-radius sample of clusters is equivalent to a sample with richness cut between R = 0 and R = 1 Abell richness classes.

With the above richness threshold, 97 clusters were selected in total. These are listed in Table 5, along with the associated redshift information. Of these, 65 have redshifts from the EM survey, including 6 that were rejected as projection effects. A further 19 cluster redshifts are available from the literature and unpublished sources. These are also listed in Table 5. With the projection effects removed, the final redshift compilation has a completeness of 86% (78/91). Table 5 is almost identical to the cluster sample used for the correlation analysis (Nichol et al. 1992) although there are some differences. We have updated the list with new redshift information and re-classified E448 as a phantom cluster. In addition, clusters E450 and E297 were upgraded from phantom clusters to genuine cluster status on the basis of new data. For completeness, in Table 5 we include also the position angles and eccentricities derived for the clusters in Martin et al. (1995). The reader is directed to this paper for a full discussion of the methods used in deriving these values.

## 8 CONCLUSIONS

We present the redshifts of the galaxies and clusters which comprise the Edinburgh–Milano cluster redshift survey. The clusters in this work were selected from the digitised survey of the EDCC and $\simeq 10$ redshifts have been secured in each cluster. The redshift data presented here for 94 clusters thus represent a large and homogeneous database. Using the galaxy redshifts we have developed an algorithm to exclude phantom clusters and interlopers from the survey.

About 10% of clusters in our survey are heavily contaminated confirming previous results that projection effects are a serious problem for optically selected rich clusters. This is the only digitised cluster survey to really attack the problem of projection effects by trying to de–project in 3–D. For 37 clusters we present line–of–sight velocity dispersions. From a comparison with structure formation models, the velocity dispersion data are most consistent with an $\Omega = 1$ CDM model with a biasing parameter $b = 1.6 - 2$ and are inconsistent with the model if $2.5 \leq b \leq 1.3$.

**Acknowledgments**

We are indebted to the supporting staff of the European Southern Observatory and the Anglo–Australian Observatory for their excellent assistance during the observations of this project. We also would like to thank N. Heydon–Dumbleton for his important contribution to the starting of this project, and J. Huchra, H. Andernach, A. Broadbent, D. Nicholson, D. Lambas, Q. Parker, J. Peacock, T. Shanks for providing both published and unpublished redshifts. Heinz Andernach is warmly acknowledged for useful suggestions to a draft of this paper. CAC acknowledges the SERC for the award of an Advanced Fellowship and RCN also acknowledges the SERC for receipt of a studentship and travel grants. This work has received partial financial support from the European Community (EEC contract ERB-CHR-XCT-920033).

**Figure Captions**

**Figure 1.** Example of a good S/N galaxy spectrum observed with EFOSC and the 3.6 m ESO telescope.

**Figure 2.** Four examples of observed clusters, chosen to show both the results of using different instrumentation and the differences in the degree of foreground/background contamination. The sky plots (left panels for E460 and E495) have a size which corresponds to the angular Abell radius of the cluster. The size of the circles is proportional to the $b_j$ magnitude of the galaxies (extracted from the EDSGC), with the small dots representing objects fainter than $b_j = 19.5$, and the largest circles (in the left panel) corresponding to galaxies brighter than $b_j = 15$. The crosses mark the objects for which a redshift appears in Table 3. E127 and E712 have been observed only at the AAT — using AUTOFIB — as indicated by the sparse distribution of the crosses: E127 is a case classified as projection on the basis of the available redshifts. E460 has been observed both at ESO and at the AAT, while E495 at ESO only. For these two cases, the bulk of redshifts is concentrated in the center of the cluster core, due to the small field of EFOSC. This region (dashed square) is zoomed in the right panel for more clarity.

**Figure 3.** Right ascension cone diagram for both the EM cluster redshifts and those found in the literature for the small–Abell–radius sample of Table 5. The declination slice runs from $-42°.5 \leq \text{Dec} \leq -22°.5$

**Figure 4.** The distribution of velocity dispersions for a sample of 37 EM clusters with $N_{clus} \geq 6$.

| Name | RA | DEC | $m_B$ | cz |
|---|---|---|---|---|
| N5740 | 14 41 53 | 01 53 | 12.5 | $1575 \pm 20$ |
| N5746 | 14 42 23 | 02 10 | 11.5 | $1801 \pm 33$ |
| N5921 | 15 19 30 | 05 15 | 12.0 | $1480 \pm 10$ |
| N6070 | 16 07 24 | 00 50 | 12.5 | $2005 \pm 7$ |
| N6118 | 16 19 12 | -02 10 | 12.0 | $1578 \pm 15$ |
| N6958 | 20 45 30 | -38 11 | 12.2 | $2742 \pm 50$ |
| N7793 | 23 57 48 | -32 35 | 9.1 | $231 \pm 7$ |
| A4038 | 23 45 08 | -28 25 | 13.7 | $8813 \pm 65$ |
| HD171391 | 18 34 37 | 10 59 | 5.1 | $6.9 \pm 0.2$ |
| HD35410 | 05 21 56 | -00 56 | 5.2 | $20.5 \pm 1.0$ |

Table 1: Characteristics of the 10 templates used with the cross–correlation routine. The last two are stellar spectra (from Parker *et al.* 1987). The others are new galaxy templates specifically observed during the survey. Literature heliocentric redshifts, in $\mathrm{km\,s^{-1}}$, are from Da Costa *et al.* (1984) .

| EDCC | RA | DEC | $b_j$ | $cz_1$ | Tel. | $cz_2$ | Tel. | $|\Delta v|$ |
|---|---|---|---|---|---|---|---|---|
| 235 | 22 59 54.3 | -33 37 11 | 17.00 | 26222 | ESO 3.6m | 26508 | ESO 3.6m | 286 |
| 311 | 23 28 38.4 | -36 44 49 | 19.22 | 28588 | ESO 3.6m | 28595 | ESO 3.6m | 7 |
| 326 | 23 35 49.9 | -38 29 55 | 18.94 | 32770 | ESO 3.6m | 32875 | ESO 2.2m | 105 |
| 326 | 23 35 57.2 | -38 30 36 | 18.15 | 19597 | ESO 3.6m | 18431 | ESO 2.2m | 1166 |
| 348 | 23 44 53.2 | -28 23 11 | 14.22 | 7989 | ESO 2.2m | 7962 | AAT | 27 |
| 348 | 23 45 07.9 | -28 25 01 | 13.82 | 8579 | ESO 2.2m | 8496 | AAT | 83 |
| 408 | 00 07 23.9 | -35 57 11 | 19.51 | 35218 | ESO 3.6m | 36573 | ESO 3.6m | 1355 |
| 408 | 00 07 24.6 | -35 57 33 | 19.09 | 36599 | ESO 3.6m | 36606 | ESO 3.6m | 7 |
| 408 | 00 07 25.7 | -35 57 08 | 18.91 | 36787 | ESO 3.6m | 37568 | ESO 3.6m | 781 |
| 408 | 00 07 25.9 | -35 58 38 | 18.97 | 37129 | ESO 3.6m | 36965 | ESO 3.6m | 164 |
| 410 | 00 08 49.2 | -29 07 58 | 15.03 | 18269 | ESO 3.6m | 18284 | ESO 2.2m | 15 |
| 448 | 00 26 21.5 | -30 22 29 | 17.82 | 30752 | ESO 3.6m | 30982 | ESO 3.6m | 230 |
| 460 | 00 35 02.0 | -28 48 14 | 16.91 | 33707 | ESO 3.6m | 33398 | ESO 3.6m | 309 |
| 460 | 00 35 02.0 | -28 48 14 | 16.91 | 33398 | ESO 3.6m | 33485 | ESO 3.6m | 87 |
| 460 | 00 35 02.0 | -28 48 14 | 16.91 | 33707 | ESO 3.6m | 33356 | AAT | 351 |
| 460 | 00 35 04.6 | -28 48 27 | 17.02 | 33267 | ESO 3.6m | 33788 | AAT | 519 |
| 460 | 00 35 09.6 | -28 49 45 | 18.53 | 34116 | ESO 3.6m | 35602 | ESO 3.6m | 1485 |
| 460 | 00 35 13.3 | -28 48 14 | 17.67 | 18548 | ESO 3.6m | 18749 | ESO 3.6m | 201 |
| 460 | 00 35 13.3 | -28 48 14 | 17.67 | 18548 | ESO 3.6m | 18476 | AAT | 72 |
| 462 | 00 34 48.4 | -39 25 16 | 18.70 | 18707 | ESO 3.6m | 18983 | ESO 3.6m | 276 |
| 462 | 00 34 58.8 | -39 24 48 | 17.29 | 19093 | ESO 3.6m | 18832 | ESO 3.6m | 261 |
| 462 | 00 35 00.2 | -39 23 34 | 17.26 | 19605 | ESO 3.6m | 19006 | ESO 3.6m | 599 |
| 462 | 00 35 02.5 | -39 24 16 | 15.82 | 18863 | ESO 3.6m | 19015 | ESO 3.6m | 152 |
| 482 | 00 47 03.5 | -29 48 05 | 18.77 | 33918 | ESO 3.6m | 33125 | AAT | 793 |
| 571 | 01 30 03.4 | -31 20 57 | 15.65 | 21372 | ESO 2.2m | 21336 | ESO 3.6m | 36 |
| 606 | 01 58 29.3 | -33 13 36 | 18.72 | 28771 | ESO 3.6m | 28510 | ESO 3.6m | 261 |
| 712 | 02 54 24.8 | -24 54 36 | 16.35 | 32936 | ESO 2.2m | 33133 | AAT | 197 |
| 658 | 02 27 49.0 | -33 23 57 | 16.28 | 23868 | ESO 2.2m | 23827 | AAT | 41 |
| 735 | 03 09 16.4 | -27 07 10 | 15.96 | 20475 | ESO 3.6m | 20412 | ESO 3.6m | 63 |
| 742 | 03 11 48.8 | -38 29 11 | 15.21 | 25930 | ESO 3.6m | 25611 | ESO 2.2m | 319 |
| 758 | 03 20 30.5 | -41 31 31 | 18.34 | 19387 | ESO 3.6m | 19390 | ESO 3.6m | 3 |

Table 2: Results for galaxies in the survey with two independent observations either at the ESO 3.6 m and 2.2 m telescopes or the AAT. Radial velocities in columns 5 and 7 are heliocentric, in units of $\mathrm{km\, s^{-1}}$.

Table 3

| EDCC | RA | DEC | $b_j$ | cz | $\Delta$ cz | Notes |
|---|---|---|---|---|---|---|
| 5 | 21 29 11.4 | -22 45 35 | 17.7 | 33895 | 114 | E |
| 5 | 21 29 13.6 | -22 45 58 | 19.4 | 38364 | 359 | E |
| 5 | 21 29 14.6 | -22 47 12 | 18.8 | 32735 | 171 | E |
| 5 | 21 29 14.6 | -22 47 37 | 19.4 | 33458 | 120 | E |
| 5 | 21 29 15.2 | -22 46 37 | 17.6 | 34675 | 659 | E |
| 5 | 21 29 15.8 | -22 48 25 | 20.4 | 41266 | 554 | E |
| 42 | 21 46 23.9 | -30 57 04 | 20.5 | 35603 | 125 | E |
| 42 | 21 46 23.3 | -30 56 27 | 18.1 | 35192 | 83 | E |
| 42 | 21 46 24.7 | -30 57 56 | 19.1 | 35189 | 77 | E |
| 42 | 21 46 30.0 | -30 57 28 | 18.6 | 35186 | 137 | E |
| 42 | 21 46 26.2 | -30 58 53 | 18.5 | 37144 | 77 | E |
| 42 | 21 46 31.6 | -30 59 33 | 20.3 | 36265 | 338 | E |
| 42 | 21 46 26.4 | -30 56 47 | 17.6 | 36178 | 56 | E |
| 42 | 21 46 29.2 | -30 58 23 | 19.6 | 32857 | 503 | E |
| 51 | 21 49 27.7 | -29 06 40 | 16.4 | 28159 | 44 | E |
| 51 | 21 49 25.1 | -29 09 32 | 17.8 | 27400 | 41 | E |
| 51 | 21 49 28.4 | -29 09 58 | 17.8 | 27931 | 65 | E |
| 51 | 21 49 23.4 | -29 00 34 | 17.0 | 27673 | 38 | E |
| 51 | 21 49 26.7 | -29 07 29 | 16.8 | 27275 | 32 | E |
| 51 | 21 49 32.6 | -29 13 18 | 16.5 | 6919 | 74 | E |
| 57 | 21 54 03.4 | -30 20 57 | 20.3 | 27553 | 56 | E |
| 57 | 21 54 13.3 | -30 20 00 | 18.9 | 27733 | 38 | E |
| 57 | 21 54 04.0 | -30 19 04 | 17.9 | 28018 | 47 | E |
| 57 | 21 53 56.5 | -30 19 06 | 17.5 | 27706 | 65 | E |
| 80 | 21 59 03.2 | -22 39 09 | 16.5 | 21716 | 71 | E |
| 80 | 21 59 04.5 | -22 39 11 | 16.7 | 21489 | 59 | E |
| 80 | 21 59 09.4 | -22 38 23 | 18.0 | 21947 | 56 | E |
| 80 | 21 59 15.6 | -22 38 13 | 17.7 | 21291 | 56 | E |
| 80 | 21 59 03.7 | -22 43 36 | 14.8 | 21270 | 59 | E |
| 80 | 22 59 59.7 | -22 42 52 | 15.0 | 21108 | 47 | E,AC |
| 80 | 21 59 01.5 | -22 39 46 | 19.0 | 20143 | 56 | E |
| 80 | 21 59 06.1 | -22 40 29 | 15.0 | 19924 | 38 | E |
| 80 | 21 59 09.9 | -22 39 27 | 16.1 | 20502 | 35 | E |
| 80 | 21 59 08.7 | -22 38 38 | 19.0 | 20280 | 65 | E |
| 80 | 21 59 08.9 | -22 37 36 | 17.6 | 19183 | 383 | E |
| 80 | 22 00 14.0 | -22 57 22 | 18.6 | 20460 | 290 | A |
| 80 | 22 00 20.3 | -23 02 35 | 17.8 | 21956 | 89 | A |
| 80 | 22 00 23.6 | -23 07 14 | 17.6 | 20466 | 62 | A |
| 80 | 22 01 05.2 | -23 11 26 | 17.4 | 20613 | 155 | A |
| 80 | 22 00 30.0 | -23 08 23 | 18.4 | 21042 | 68 | A |
| 80 | 22 00 08.0 | -23 13 33 | 17.5 | 21297 | 89 | A |
| 80 | 21 59 37.5 | -23 11 39 | 16.6 | 21048 | 53 | A |
| 80 | 21 59 35.9 | -23 09 26 | 17.2 | 21564 | 56 | A |
| 80 | 21 59 43.8 | -23 07 56 | 17.1 | 20937 | 100 | A |
| 80 | 21 59 37.8 | -23 07 41 | 17.2 | 21516 | 51 | A |
| 80 | 21 59 49.9 | -23 01 36 | 18.1 | 21830 | 44 | A |
| 80 | 21 59 40.3 | -23 02 23 | 16.6 | 19357 | 35 | A |
| 80 | 21 58 47.7 | -22 59 11 | 16.9 | 21450 | 101 | A |
| 80 | 21 58 43.2 | -22 56 52 | 17.7 | 19177 | 122 | A |
| 80 | 21 59 10.6 | -22 53 46 | 17.1 | 21057 | 98 | A |
| 80 | 21 59 18.9 | -22 51 00 | 18.5 | 21812 | 80 | A |
| 80 | 21 59 36.9 | -22 46 32 | 17.0 | 21791 | 122 | A |
| 80 | 21 59 47.3 | -22 58 00 | 18.3 | 22217 | 251 | A |
| 80 | 21 58 57.4 | -23 11 24 | 17.3 | 32833 | 77 | A |
| 80 | 21 59 38.8 | -22 56 03 | 18.8 | 10189 | 431 | A |
| 80 | 21 59 58.8 | -22 58 32 | 18.3 | 32680 | 104 | A |
| 80 | 22 00 08.9 | -22 52 53 | 18.1 | 13847 | 86 | A |
| 80 | 22 00 24.3 | -23 03 23 | 16.8 | 16200 | 71 | A |
| 80 | 22 00 44.0 | -22 57 40 | 17.6 | 22148 | 57 | A |
| 80 | 22 00 07.4 | -22 58 26 | 17.6 | 22051 | 75 | A |
| 99 | 22 06 30.2 | -27 33 09 | 17.8 | 19105 | 38 | E |
| 99 | 22 06 32.9 | -27 33 30 | 17.6 | 19135 | 41 | E |
| 99 | 22 06 23.1 | -27 34 29 | 17.2 | 27140 | 38 | E |
| 99 | 22 06 24.3 | -27 33 58 | 18.9 | 26507 | 101 | E |
| 114 | 22 11 08.8 | -36 55 44 | 15.3 | 10180 | 68 | E |
| 114 | 22 11 15.4 | -36 56 37 | 16.0 | 10183 | 53 | E |
| 115 | 22 10 24.2 | -34 54 51 | 19.0 | 21821 | 294 | E |
| 115 | 22 10 34.2 | -34 55 01 | 17.8 | 21995 | 56 | E |
| 115 | 22 10 38.7 | -34 54 29 | 18.2 | 21833 | 65 | E |
| 115 | 22 10 36.8 | -34 54 36 | 18.2 | 45229 | 65 | E |

| EDCC | RA | DEC | $b_j$ | cz | $\Delta$ cz | Notes |
|---|---|---|---|---|---|---|
| 124 | 22 14 41.2 | -35 58 00 | 18.0 | 43511 | 104 | E |
| 124 | 22 14 42.1 | -35 57 57 | 19.4 | 42102 | 38 | E |
| 124 | 22 14 43.4 | -35 58 08 | 20.1 | 44219 | 197 | E |
| 124 | 22 14 47.8 | -36 00 07 | 19.7 | 44489 | 62 | E |
| 124 | 22 14 57.6 | -35 57 57 | 20.8 | 45538 | 95 | E,AC |
| 124 | 22 14 51.6 | -35 58 42 | 18.0 | 43646 | 131 | E |
| 124 | 22 14 53.2 | -35 58 50 | 18.3 | 43643 | 47 | E |
| 124 | 22 14 54.1 | -35 59 17 | 18.4 | 44462 | 77 | E |
| 124 | 22 14 48.2 | -35 58 07 | 19.0 | 48104 | 47 | E |
| 124 | 22 14 55.5 | -35 57 58 | 20.1 | 46066 | 65 | E,AC |
| 127 | 22 15 40.5 | -39 08 45 | 18.5 | 41949 | 56 | A |
| 127 | 22 15 52.6 | -39 12 54 | 18.0 | 41431 | 89 | A |
| 127 | 22 15 37.4 | -39 08 34 | 18.9 | 41032 | 122 | A |
| 127 | 22 15 31.1 | -39 00 07 | 18.8 | 41077 | 140 | A,AC |
| 127 | 22 16 54.0 | -39 06 00 | 17.5 | 11985 | 41 | A |
| 127 | 22 15 55.1 | -39 19 50 | 18.0 | 13229 | 95 | A |
| 127 | 22 15 18.6 | -39 08 36 | 17.3 | 14237 | 83 | A |
| 127 | 22 14 23.4 | -39 08 08 | 18.2 | 11880 | 321 | A |
| 127 | 22 15 17.7 | -39 13 09 | 17.7 | 16347 | 71 | A |
| 127 | 22 14 49.3 | -39 13 51 | 18.8 | 21498 | 311 | A |
| 127 | 22 14 07.5 | -39 00 37 | 18.4 | 19096 | 307 | A |
| 127 | 22 16 40.3 | -39 16 05 | 17.0 | 25422 | 59 | A |
| 127 | 22 16 19.8 | -39 17 25 | 17.1 | 28261 | 131 | A |
| 127 | 22 16 08.7 | -39 15 51 | 18.4 | 44845 | 86 | A |
| 131 | 22 17 06.0 | -34 51 02 | 17.9 | 47493 | 104 | E |
| 131 | 22 16 46.2 | -34 50 02 | 17.5 | 46749 | 26 | E,AC |
| 131 | 22 16 57.8 | -34 49 57 | 18.4 | 47058 | 53 | E |
| 131 | 22 17 07.8 | -34 52 19 | 18.0 | 46230 | 62 | E |
| 145 | 22 25 01.0 | -30 49 04 | 17.1 | 16788 | 50 | A |
| 145 | 22 25 05.2 | -30 48 55 | 17.3 | 18934 | 50 | A |
| 145 | 22 26 04.5 | -30 53 06 | 17.8 | 17876 | 53 | A |
| 145 | 22 25 30.3 | -30 53 34 | 17.3 | 16974 | 65 | A |
| 145 | 22 25 44.4 | -30 56 20 | 17.8 | 17118 | 56 | A |
| 145 | 22 25 16.1 | -30 52 47 | 17.0 | 15382 | 134 | A |
| 145 | 22 25 03.9 | -30 53 16 | 17.2 | 17807 | 56 | A |
| 145 | 22 25 18.4 | -31 01 15 | 17.6 | 16659 | 44 | A |
| 145 | 22 25 07.0 | -30 57 15 | 16.5 | 17672 | 59 | A |
| 145 | 22 24 56.0 | -30 59 28 | 16.7 | 14767 | 59 | A |
| 145 | 22 24 34.2 | -30 45 03 | 16.2 | 17576 | 107 | A |
| 145 | 22 25 28.1 | -30 40 58 | 18.1 | 7701 | 38 | A |
| 172 | 22 36 15.4 | -36 51 06 | 16.2 | 17070 | 32 | E |
| 172 | 22 35 44.0 | -37 01 33 | 17.3 | 17501 | 38 | E |
| 172 | 22 35 54.1 | -37 00 18 | 15.8 | 18050 | 38 | E |
| 172 | 22 36 04.6 | -36 53 16 | 16.7 | 23209 | 47 | E |
| 175 | 22 36 47.4 | -38 07 06 | 17.9 | 46123 | 41 | E |
| 175 | 22 36 44.5 | -38 07 18 | 18.1 | 46284 | 32 | E |
| 175 | 22 36 59.8 | -38 05 56 | 17.6 | 45682 | 38 | E |
| 175 | 22 35 55.3 | -38 05 19 | 18.3 | 56915 | 38 | E |
| 178 | - | - | - | 14824 | 68 | E,NC |
| 178 | - | - | - | 14983 | 38 | E,NC |
| 198 | 22 46 35.1 | -41 10 48 | 18.6 | 37989 | 41 | E |
| 198 | 22 46 33.0 | -41 10 33 | 18.7 | 37536 | 47 | E |
| 198 | 22 46 37.5 | -41 09 27 | 18.8 | 37863 | 35 | E |
| 198 | 22 46 23.1 | -41 10 52 | 19.1 | 57188 | 38 | E |
| 198 | 22 46 22.0 | -41 11 04 | 19.2 | 58129 | 59 | E |
| 198 | 22 46 43.2 | -41 07 39 | 19.4 | 62887 | 98 | E |
| 201 | 22 46 48.4 | -31 32 00 | 17.2 | 31202 | 26 | E |
| 201 | 22 46 48.8 | -31 30 57 | 17.6 | 32026 | 29 | E |
| 201 | 22 47 10.8 | -31 23 59 | 18.1 | 34598 | 47 | E |
| 201 | 22 47 17.6 | -31 23 30 | 17.1 | 33891 | 35 | E |
| 216 | 22 50 48.2 | -25 49 20 | 16.9 | 23269 | 29 | E |
| 216 | 22 50 54.7 | -25 47 29 | 19.9 | 25869 | 59 | E |
| 216 | 22 50 46.4 | -25 49 03 | 18.7 | 26978 | 56 | E |
| 216 | 22 50 50.6 | -25 48 45 | 20.6 | 46773 | 50 | E |
| 216 | 22 50 51.3 | -25 48 16 | 20.0 | 45439 | 38 | E |
| 230 | 22 55 59.9 | -31 08 29 | 17.8 | 33525 | 41 | E |
| 230 | 22 55 58.5 | -31 09 02 | 18.3 | 32302 | 32 | E |
| 230 | 22 55 34.4 | -31 07 01 | 17.2 | 24391 | 65 | E |
| 230 | 22 55 37.3 | -31 06 50 | 18.3 | 37821 | 47 | E |

| EDCC | RA | DEC | $b_j$ | cz | $\Delta$ cz | Notes |
|---|---|---|---|---|---|---|
| 235 | 22 59 49.4 | -33 37 09 | 18.6 | 26183 | 80 | E |
| 235 | 22 59 54.3 | -33 37 11 | 17.2 | 26222 | 35 | E |
| 235 | 22 59 49.3 | -33 36 27 | 19.4 | 19354 | 137 | E |
| 235 | 23 00 06.9 | -33 39 32 | 17.2 | 19720 | 26 | E |
| 247 | 23 02 59.4 | -39 21 03 | 18.4 | 49840 | 53 | E |
| 247 | 23 02 55.0 | -39 21 38 | 18.1 | 49720 | 56 | E |
| 247 | 23 02 58.0 | -39 20 08 | 18.7 | 26408 | 41 | E |
| 247 | 23 02 58.1 | -39 18 50 | 18.4 | 50781 | 44 | E |
| 256 | 23 07 52.8 | -23 12 20 | 18.3 | 27359 | 32 | E |
| 256 | 23 07 46.2 | -23 12 06 | 17.9 | 27143 | 26 | E |
| 256 | 23 07 50.6 | -23 18 39 | 18.1 | 32824 | 32 | E |
| 256 | 23 07 58.0 | -23 17 44 | 18.0 | 33079 | 29 | E |
| 261 | 23 08 59.1 | -29 22 03 | 17.4 | 25842 | 41 | E |
| 261 | 23 08 53.3 | -29 21 32 | 17.8 | 25893 | 26 | E |
| 261 | 23 09 01.9 | -29 23 34 | 18.1 | 36640 | 29 | E |
| 261 | 23 08 57.4 | -29 22 58 | 17.1 | 35036 | 80 | E |
| 261 | 23 08 56.0 | -29 21 41 | 18.4 | 36697 | 95 | E |
| 261 | - | - | - | 34344 | 71 | E,NC |
| 261 | - | - | - | 35588 | 56 | E,NC |
| 261 | - | - | - | 35381 | 65 | E,NC |
| 269 | 23 11 47.9 | -38 01 44 | 18.3 | 28150 | 47 | A |
| 269 | 23 11 33.8 | -38 00 00 | 17.7 | 28021 | 29 | A |
| 269 | 23 12 13.6 | -38 02 00 | 18.6 | 27574 | 59 | A |
| 269 | 23 12 05.4 | -38 00 26 | 17.0 | 28105 | 74 | A |
| 269 | 23 11 40.6 | -37 49 39 | 18.0 | 28225 | 38 | A |
| 269 | 23 13 53.8 | -38 12 19 | 17.6 | 18898 | 38 | A |
| 269 | 23 12 57.1 | -38 12 04 | 17.6 | 20092 | 35 | A |
| 269 | 23 12 10.2 | -38 19 45 | 17.0 | 18137 | 29 | A |
| 269 | 23 11 52.5 | -37 58 10 | 19.0 | 20274 | 83 | A |
| 269 | 23 12 52.9 | -38 00 10 | 18.0 | 53635 | 83 | A |
| 269 | 23 13 01.9 | -38 04 09 | 17.9 | 53321 | 29 | A |
| 269 | 23 12 18.0 | -38 03 17 | 19.0 | 55497 | 74 | A |
| 285 | 23 17 49.4 | -42 03 53 | 17.9 | 17133 | 29 | E |
| 285 | 23 17 49.2 | -42 04 30 | 17.0 | 16518 | 35 | E,AC |
| 285 | 23 19 08.3 | -42 10 47 | 17.7 | 26090 | 41 | E |
| 285 | 23 18 56.6 | -42 10 12 | 16.0 | 26933 | 44 | E,AC |

| EDCC | RA | DEC | $b_j$ | cz | $\Delta$ cz | Notes |
|---|---|---|---|---|---|---|
| 297 | 23 24 05.8 | -24 09 36 | 18.9 | 26642 | 74 | E |
| 297 | 23 23 55.8 | -24 08 38 | 19.0 | 27113 | 44 | E |
| 297 | 23 24 03.1 | -24 07 45 | 17.2 | 26498 | 35 | E |
| 297 | 23 24 08.4 | -24 09 15 | 19.9 | 32971 | 419 | E |
| 297 | 23 23 53.3 | -24 09 09 | 19.8 | 35240 | 68 | E |
| 307 | 23 26 58.9 | -39 41 51 | 16.9 | 16188 | 47 | E |
| 307 | 23 27 19.6 | -39 37 11 | 17.1 | 15945 | 32 | E |
| 307 | 23 27 15.5 | -39 43 39 | 17.1 | 16968 | 38 | E |
| 311 | 23 28 36.4 | -36 47 15 | 19.7 | 28819 | 35 | E |
| 311 | 23 28 38.4 | -36 44 49 | 19.2 | 28588 | 119 | E |
| 311 | 23 28 41.1 | -36 46 42 | 19.2 | 28579 | 38 | E |
| 311 | 23 28 45.6 | -36 46 16 | 17.9 | 28486 | 161 | E |
| 311 | 23 28 31.4 | -36 47 29 | 18.6 | 30311 | 86 | E |
| 311 | 23 28 37.2 | -36 47 07 | 18.2 | 29664 | 98 | E |
| 311 | 23 28 44.0 | -36 45 33 | 20.2 | 26957 | 41 | E |
| 316 | 23 30 07.8 | -36 28 01 | 18.2 | 27940 | 47 | E |
| 316 | 23 30 10.1 | -36 31 05 | 17.9 | 28270 | 35 | E |
| 316 | 23 30 11.1 | -36 30 28 | 17.3 | 28549 | 41 | E |
| 316 | 23 30 08.1 | -36 28 58 | 17.2 | 28672 | 47 | E |
| 326 | 23 35 43.9 | -38 29 45 | 18.8 | 32389 | 59 | E |
| 326 | 23 35 49.2 | -38 31 27 | 19.2 | 32590 | 35 | E |
| 326 | 23 35 49.9 | -38 29 55 | 18.9 | 32770 | 53 | E |
| 326 | 23 35 57.5 | -38 30 36 | 18.2 | 18431 | 59 | E |
| 326 | 23 35 49.9 | -38 30 08 | 20.0 | 49531 | 104 | E,AC |
| 326 | 23 35 53.6 | -38 30 08 | 18.9 | 49702 | 68 | E |
| 326 | 23 35 57.5 | -38 30 11 | 19.6 | 71155 | 59 | E |
| 332 | 23 38 58.4 | -29 30 50 | 15.5 | 15217 | 44 | E |
| 332 | 23 38 44.2 | -29 25 52 | 18.9 | 15661 | 83 | E |
| 332 | 23 38 52.8 | -29 35 53 | 15.1 | 15670 | 38 | E |
| 348 | 23 45 09.9 | -28 23 08 | 18.2 | 8064 | 26 | A |
| 348 | 23 46 08.0 | -28 25 17 | 14.8 | 7791 | 89 | A,AC |
| 348 | 23 45 45.3 | -28 23 39 | 16.0 | 8864 | 26 | A |
| 348 | 23 45 23.1 | -28 25 11 | 16.7 | 7359 | 14 | A |
| 348 | 23 45 50.4 | -28 33 24 | 16.6 | 8130 | 20 | A |

| EDCC | RA | DEC | $b_j$ | cz | $\Delta$ cz | Notes |
|---|---|---|---|---|---|---|
| 348 | 23 45 18.3 | -28 26 36 | 17.5 | 9002 | 14 | A |
| 348 | 23 45 13.5 | -28 34 09 | 18.1 | 8864 | 23 | A |
| 348 | 23 45 06.8 | -28 27 04 | 18.3 | 8241 | 26 | A,AC |
| 348 | 23 44 53.2 | -28 23 11 | 14.1 | 7962 | 32 | A |
| 348 | 23 44 35.4 | -28 31 06 | 18.3 | 7701 | 29 | A |
| 348 | 23 45 06.8 | -28 26 04 | 18.3 | 7911 | 29 | A |
| 348 | 23 44 38.4 | -28 28 15 | 17.0 | 8756 | 20 | A |
| 348 | 23 45 07.9 | -28 25 01 | 13.8 | 8496 | 23 | A |
| 348 | 23 44 40.9 | -28 24 07 | 16.7 | 8804 | 20 | A |
| 348 | 23 44 59.6 | -28 24 12 | 17.1 | 9185 | 17 | A |
| 348 | 23 44 53.2 | -28 23 11 | 14.1 | 8112 | 86 | A |
| 348 | 23 45 04.0 | -28 22 52 | 18.4 | 9161 | 23 | A |
| 348 | 23 44 44.2 | -28 20 27 | 17.4 | 8295 | 29 | A |
| 348 | 23 44 38.5 | -28 18 29 | 18.9 | 8223 | 44 | A |
| 348 | 23 44 54.7 | -28 19 16 | 17.6 | 8163 | 29 | A |
| 348 | 23 44 46.0 | -28 15 12 | 17.1 | 7317 | 14 | A |
| 348 | 23 45 03.5 | -28 17 26 | 17.5 | 7833 | 23 | A |
| 348 | 23 45 23.1 | -28 25 11 | 16.7 | 9596 | 14 | A |
| 348 | 23 45 43.3 | -28 27 25 | 16.8 | 10486 | 20 | A |
| 348 | 23 45 25.5 | -28 26 28 | 18.1 | 9599 | 20 | A |
| 348 | 23 45 13.7 | -28 28 54 | 17.4 | 9545 | 17 | A |
| 348 | 23 45 14.5 | -28 25 49 | 17.8 | 9773 | 29 | A |
| 348 | 23 44 52.2 | -28 32 59 | 17.4 | 10102 | 29 | A |
| 348 | 23 44 58.8 | -28 25 56 | 18.0 | 9599 | 20 | A |
| 348 | 23 44 47.3 | -28 23 49 | 16.5 | 9647 | 17 | A |
| 348 | 23 44 37.4 | -28 17 42 | 17.0 | 9971 | 23 | A |
| 348 | 23 43 04.5 | -28 16 47 | 18.1 | 9641 | 29 | A |
| 348 | 23 44 37.4 | -28 29 27 | 18.4 | 18940 | 26 | A |
| 366 | 23 52 11.5 | -27 58 04 | 19.1 | 22580 | 80 | E |
| 366 | 23 52 10.8 | -27 57 35 | 16.9 | 21836 | 47 | E |
| 366 | 23 52 10.9 | -27 56 34 | 18.2 | 21686 | 53 | E |
| 366 | 23 52 10.7 | -27 56 22 | 18.2 | 20499 | 332 | E |
| 366 | 23 52 12.7 | -27 55 54 | 17.8 | 21426 | 74 | E |
| 366 | 23 52 14.5 | -27 55 50 | 17.8 | 22016 | 29 | E |
| 366 | 23 52 14.5 | -27 55 26 | 18.4 | 22691 | 65 | E |
| 366 | 23 52 18.8 | -27 57 34 | 18.4 | 15760 | 95 | E |
| 372 | 23 54 26.0 | -34 57 32 | 14.7 | 12675 | 35 | E |
| 392 | 00 00 02.4 | -34 58 45 | 19.2 | 33888 | 295 | E |
| 392 | 00 00 03.8 | -34 58 02 | 19.5 | 34419 | 71 | E |
| 392 | 00 00 12.4 | -34 57 21 | 18.0 | 33402 | 62 | E |
| 392 | 00 00 10.7 | -34 55 37 | 18.7 | 33327 | 53 | E |
| 392 | 00 00 16.5 | -34 56 15 | 18.5 | 33927 | 53 | E |
| 392 | 00 00 06.4 | -34 58 18 | 19.0 | 33127 | 56 | E |
| 392 | 00 00 10.5 | -34 57 19 | 18.2 | 32650 | 41 | E |
| 392 | 00 00 13.3 | -34 58 25 | 18.1 | 35873 | 41 | E |
| 394 | 00 00 07.4 | -36 12 51 | 17.8 | 14950 | 59 | E |
| 394 | 00 00 10.5 | -36 12 28 | 17.8 | 14668 | 29 | E |
| 394 | 00 00 15.6 | -36 11 44 | 18.2 | 14431 | 38 | E |
| 394 | 00 00 39.5 | -36 12 55 | 14.6 | 14728 | 47 | E |
| 394 | 00 00 11.6 | -36 12 31 | 19.1 | 13808 | 56 | E |
| 394 | 00 00 17.2 | -36 12 07 | 18.9 | 15724 | 77 | E |
| 400 | 00 03 39.1 | -34 59 45 | 19.4 | 34077 | 107 | E |
| 400 | 00 03 43.3 | -34 59 06 | 19.3 | 32452 | 80 | E |
| 400 | 00 03 46.4 | -34 57 34 | 19.0 | 35453 | 92 | E |
| 400 | 00 03 42.7 | -34 56 32 | 20.4 | 34619 | 119 | E |
| 400 | 00 03 44.6 | -34 56 52 | 19.9 | 32878 | 146 | E |
| 400 | 00 03 41.8 | -34 59 19 | 19.2 | 17151 | 71 | E |
| 400 | 00 03 42.8 | -34 57 51 | 18.3 | 35987 | 44 | E |
| 400 | 00 03 43.2 | -34 58 30 | 18.5 | 17765 | 86 | E |
| 400 | 00 03 46.0 | -35 00 32 | 19.0 | 86762 | 68 | E |
| 400 | 00 03 46.8 | -34 58 01 | 19.4 | 35576 | 146 | E |
| 408 | 00 07 23.9 | -35 57 11 | 19.5 | 36573 | 143 | E |
| 408 | 00 07 25.0 | -35 57 20 | 20.5 | 35069 | 113 | E,AC |
| 408 | 00 07 26.0 | -35 56 49 | 21. | 34784 | 53 | E,AC |
| 408 | 00 07 26.8 | -35 56 33 | 17.4 | 35107 | 113 | E |
| 408 | 00 07 23.0 | -38 58 30 | 20.0 | 36058 | 74 | E,AC |
| 408 | 00 07 28.2 | -35 56 52 | 19.3 | 35492 | 83 | E |
| 408 | 00 07 28.9 | -35 56 41 | 17.9 | 34281 | 89 | E |
| 408 | 00 07 30.0 | -35 56 30 | 19.5 | 34748 | 74 | E,AC |
| 408 | 00 07 25.9 | -35 58 38 | 19.0 | 37129 | 50 | E |
| 408 | 00 07 25.7 | -35 57 08 | 18.9 | 36787 | 74 | E |
| 408 | 00 07 24.6 | -35 57 33 | 19.1 | 36606 | 50 | E |
| 408 | 00 07 25.7 | -35 57 08 | 18.9 | 36736 | 50 | E |
| 408 | 00 07 21.0 | -35 58 23 | 19.4 | 36841 | 53 | E |
| 408 | 00 07 24.9 | -35 57 14 | 19.7 | 37767 | 374 | E |

| EDCC | RA | DEC | $b_j$ | cz | $\Delta$ cz | Notes |
|---|---|---|---|---|---|---|
| 410 | 00 08 49.5 | -29 07 58 | 15.0 | 18269 | 32 | E |
| 410 | 00 09 11.6 | -29 12 37 | 16.0 | 18617 | 29 | E |
| | | | | | | |
| 418 | 00 13 11.6 | -23 58 42 | 17.1 | 19234 | 80 | A |
| 418 | 00 13 31.2 | -24 01 52 | 17.0 | 20361 | 53 | A |
| 418 | 00 13 26.0 | -24 04 12 | 16.7 | 18997 | 56 | A |
| 418 | 00 13 33.2 | -24 00 29 | 16.4 | 20370 | 49 | A |
| 418 | 00 12 35.0 | -24 22 23 | 15.7 | 19846 | 56 | A |
| 418 | 00 12 08.0 | -24 17 49 | 16.9 | 20811 | 62 | A |
| 418 | 00 12 46.5 | -24 08 18 | 17.4 | 19420 | 53 | A |
| 418 | 00 12 06.1 | -24 01 15 | 17.3 | 18671 | 71 | A |
| 418 | 00 12 37.0 | -24 00 34 | 17.9 | 48011 | 53 | A |
| | | | | | | |
| 419 | 00 12 31.2 | -26 32 14 | 18.1 | 39026 | 38 | A |
| 419 | 00 12 46.1 | -26 19 09 | 18.5 | 36832 | 35 | A |
| 419 | 00 11 38.3 | -26 27 24 | 17.7 | 25964 | 38 | A |
| 419 | 00 12 38.1 | -26 22 46 | 18.1 | 35786 | 74 | A |
| 419 | 00 12 50.0 | -26 21 07 | 17.8 | 43499 | 56 | A,AC |
| | | | | | | |
| 421 | 00 13 38.8 | -35 10 19 | 19.7 | 45184 | 128 | E |
| 421 | 00 13 39.7 | -35 10 23 | 18.0 | 45034 | 26 | E |
| 421 | 00 13 40.5 | -35 11 12 | 19.6 | 42411 | 170 | E |
| 421 | 00 13 38.6 | -35 12 15 | 19.9 | 42462 | 86 | E |
| 421 | 00 13 41.2 | -35 12 19 | 18.2 | 44024 | 68 | E |
| 421 | 00 13 30.3 | -35 09 59 | 19.4 | 29193 | 38 | E |
| 421 | 00 13 39.0 | -35 09 48 | 20.0 | 96281 | 575 | E,AC |
| | | | | | | |
| 424 | - | - | - | 22340 | 26 | E,NC |
| 424 | - | - | - | 22916 | 47 | E,NC |
| | | | | | | |
| 429 | 00 15 05.6 | -35 27 28 | 18.5 | 29595 | 59 | E |
| 429 | 00 15 04.4 | -35 27 40 | 16.8 | 28573 | 89 | E |
| 429 | 00 15 01.4 | -35 28 35 | 17.8 | 29523 | 38 | E |
| 429 | 00 15 09.2 | -35 27 51 | 19.1 | 27706 | 83 | E |
| 429 | 00 15 02.1 | -35 29 06 | 18.4 | 29613 | 56 | E |
| 429 | 00 15 05.7 | -35 28 49 | 18.0 | 28878 | 62 | E |
| 429 | 00 15 07.5 | -35 28 39 | 19.0 | 29550 | 188 | E |
| 429 | 00 15 15.0 | -35 29 11 | 19.0 | 30488 | 56 | E |
| 429 | 00 15 05.4 | -35 29 20 | 18.6 | 27460 | 56 | E |
| 429 | 00 15 10.3 | -35 28 40 | 16.8 | 28753 | 179 | E |

| EDCC | RA | DEC | $b_j$ | cz | $\Delta$ cz | Notes |
|---|---|---|---|---|---|---|
| 429 | 00 15 05.2 | -35 29 39 | 18.1 | 28872 | 44 | E |
| 429 | 00 15 11.6 | -35 28 47 | 18.4 | 27377 | 62 | E |
| 429 | 00 15 06.1 | -35 29 50 | 17.3 | 29328 | 47 | E |
| 429 | 00 15 05.7 | -35 30 04 | 18.5 | 29208 | 53 | E |
| 429 | 00 15 16.4 | -35 25 56 | 18.1 | 29403 | 53 | E |
| 429 | 00 15 08.1 | -35 30 25 | 18.9 | 29997 | 35 | E |
| 429 | 00 15 10.4 | -35 30 30 | 19.3 | 29676 | 80 | E |
| 429 | 00 15 08.3 | -35 27 46 | 20.3 | 19333 | 44 | E |
| 429 | 00 15 10.0 | -35 29 46 | 19.3 | 36382 | 59 | E |
| | | | | | | |
| 437 | 00 18 14.7 | -25 45 27 | 18.0 | 19936 | 179 | A |
| 437 | 00 18 11.4 | -25 49 28 | 17.9 | 19258 | 50 | A |
| 437 | 00 17 35.2 | -26 00 24 | 17.9 | 19429 | 53 | A |
| 437 | 00 17 56.0 | -25 51 33 | 18.0 | 17417 | 313 | A |
| 437 | 00 17 21.0 | -25 53 43 | 17.9 | 18943 | 35 | A |
| 437 | 00 17 06.4 | -25 54 59 | 17.5 | 33756 | 86 | A |
| | | | | | | |
| 438 | 00 20 14.9 | -38 23 38 | 17.9 | 35711 | 83 | E |
| 438 | 00 20 16.8 | -38 23 30 | 17.1 | 35750 | 44 | E |
| 438 | 00 20 11.5 | -38 24 09 | 18.5 | 36838 | 101 | E |
| 438 | 00 20 12.6 | -38 24 02 | 18.6 | 34883 | 44 | E |
| | | | | | | |
| 447 | 00 26 06.9 | -23 55 25 | 19.1 | 33216 | 71 | E,AC |
| 447 | 00 26 06.5 | -23 54 56 | 17.8 | 33564 | 26 | E,AC |
| 447 | 00 26 08.5 | -23 52 36 | 17.6 | 34371 | 26 | E |
| 447 | 00 26 11.2 | -23 52 49 | 18.1 | 34239 | 38 | E |
| 447 | 00 26 02.0 | -23 54 47 | 18.7 | 58069 | 35 | E |
| 447 | 00 26 09.4 | -23 54 34 | 19.2 | 31819 | 29 | E |
| | | | | | | |
| 448 | 00 26 14.8 | -30 23 24 | 19.2 | 30515 | 68 | E |
| 448 | 00 26 17.6 | -30 22 58 | 19.4 | 30258 | 41 | E |
| 448 | 00 26 18.7 | -30 23 06 | 19.3 | 31235 | 140 | E |
| 448 | 00 26 20.5 | -30 23 01 | 20.1 | 32506 | 116 | E |
| 448 | 00 26 21.5 | -30 22 29 | 17.8 | 30752 | 38 | E |
| 448 | 00 26 27.5 | -30 22 20 | 19.5 | 30809 | 74 | E |
| 448 | 00 26 28.7 | -30 21 39 | 18.0 | 30638 | 71 | E,AC |
| 448 | 00 26 29.0 | -30 21 40 | 18.0 | 30962 | 29 | E,AC |
| 448 | 00 25 50.2 | -30 31 20 | 18.6 | 32884 | 92 | E |
| 448 | 00 25 55.0 | -30 31 11 | 18.6 | 31178 | 83 | E |
| 448 | 00 26 17.1 | -30 22 23 | 18.5 | 38394 | 38 | E |

| EDCC | RA | DEC | $b_j$ | cz | $\Delta$ cz | Notes |
|---|---|---|---|---|---|---|
| 448 | 00 26 18.4 | -30 22 54 | 19.0 | 38139 | 38 | E |
| 448 | 00 26 19.3 | -30 22 42 | 18.8 | 38715 | 56 | E |
| 448 | 00 26 22.4 | -30 21 08 | 18.6 | 35477 | 44 | E |
| 448 | 00 25 46.8 | -30 31 12 | 18.5 | 36244 | 41 | E |
| 448 | 00 25 48.3 | -30 32 13 | 18.9 | 36103 | 65 | E |
| 448 | 00 25 44.5 | -30 29 11 | 16.8 | 21333 | 65 | E |
| 450 | 00 27 12.8 | -29 42 53 | 18.0 | 30671 | 44 | E |
| 450 | 00 27 17.8 | -29 41 34 | 18.7 | 30225 | 44 | E |
| 450 | 00 27 19.0 | -29 41 26 | 18.0 | 28690 | 44 | E |
| 450 | 00 27 24.7 | -29 42 01 | 18.3 | 28947 | 32 | E |
| 450 | 00 27 19.4 | -29 41 00 | 20.9 | 38679 | 41 | E,AC |
| 460 | 00 35 00.9 | -28 48 12 | 18.6 | 32473 | 86 | E |
| 460 | 00 35 14.2 | -28 43 37 | 19.6 | 33393 | 137 | E |
| 460 | 00 35 04.6 | -28 48 27 | 17.0 | 33267 | 41 | E |
| 460 | 00 35 06.0 | -28 48 38 | 18.0 | 32989 | 59 | E |
| 460 | 00 35 09.6 | -28 49 45 | 18.5 | 34116 | 32 | E |
| 460 | 00 34 59.2 | -28 47 24 | 20.1 | 34850 | 200 | E |
| 460 | 00 35 04.6 | -28 49 25 | 20.5 | 38730 | 68 | E |
| 460 | 00 35 05.5 | -28 48 34 | 18.4 | 32680 | 128 | E |
| 460 | 00 35 07.4 | -28 49 11 | 19.3 | 31984 | 59 | E |
| 460 | 00 35 13.4 | -28 48 14 | 17.7 | 18548 | 53 | E |
| 460 | 00 35 05.6 | -28 48 52 | 18.2 | 34023 | 35 | E |
| 460 | 00 35 06.8 | -28 47 21 | 19.7 | 32764 | 74 | E |
| 460 | 00 35 11.6 | -28 46 17 | 19.1 | 35120 | 158 | E |
| 460 | 00 35 14.7 | -28 49 43 | 18.9 | 34212 | 83 | E |
| 460 | 00 35 06.0 | -28 41 57 | 17.4 | 34308 | 56 | A |
| 460 | 00 35 02.6 | -28 48 14 | 16.8 | 33354 | 53 | A |
| 460 | 00 35 08.8 | -28 50 40 | 17.9 | 34074 | 53 | A |
| 460 | 00 34 51.1 | -28 47 17 | 18.6 | 34095 | 26 | A |
| 460 | 00 34 09.2 | -28 47 56 | 18.6 | 33852 | 50 | A |
| 460 | 00 33 35.0 | -28 52 04 | 18.2 | 33897 | 47 | A |
| 460 | 00 33 50.2 | -28 48 20 | 17.9 | 33043 | 38 | A |
| 460 | 00 33 28.4 | -28 31 40 | 18.1 | 33552 | 53 | A |
| 460 | 00 34 19.3 | -28 37 34 | 17.3 | 33756 | 41 | A |
| 460 | 00 33 40.6 | -28 36 02 | 18.5 | 7090 | 68 | A |
| 460 | 00 33 48.6 | -28 52 17 | 18.4 | 27409 | 47 | A |
| 460 | 00 33 57.9 | -28 34 29 | 16.1 | 22208 | 50 | A |
| 460 | 00 34 03.6 | -28 34 00 | 17.0 | 22250 | 47 | A |
| 460 | 00 34 04.8 | -28 32 21 | 17.9 | 22241 | 32 | A |
| 460 | 00 34 23.1 | -28 38 31 | 14.2 | 7108 | 17 | A |
| 460 | 00 34 40.7 | -28 32 43 | 16.6 | 21785 | 56 | A |
| 460 | 00 35 44.8 | -28 35 40 | 18.6 | 21785 | 26 | A |
| 462 | 00 34 48.2 | -39 25 16 | 18.7 | 18707 | 56 | E |
| 462 | 00 34 56.7 | -39 25 43 | 19.4 | 18671 | 62 | E |
| 462 | 00 34 57.4 | -39 25 35 | 18.9 | 17858 | 35 | E |
| 462 | 00 34 59.1 | -39 25 18 | 18.3 | 17807 | 23 | E |
| 462 | 00 34 52.7 | -39 24 34 | 18.5 | 19675 | 239 | E |
| 462 | 00 34 59.0 | -39 24 48 | 17.3 | 18832 | 23 | E |
| 462 | 00 35 02.5 | -39 24 16 | 15.8 | 19015 | 26 | E |
| 462 | 00 34 59.9 | -39 23 44 | 17.8 | 19669 | 29 | E |
| 462 | 00 35 00.2 | -39 23 34 | 17.3 | 19006 | 482 | E |
| 462 | 00 35 00.0 | -39 24 43 | 19.0 | 33999 | 104 | E |
| 462 | 00 34 46.3 | -39 27 17 | 17.0 | 19069 | 29 | E |
| 462 | 00 34 54.9 | -39 25 53 | 19.0 | 19180 | 41 | E |
| 470 | 00 37 16.7 | -26 29 55 | 18.2 | 33408 | 44 | E |
| 470 | 00 37 17.7 | -26 29 57 | 18.7 | 32491 | 26 | E |
| 470 | 00 37 20.9 | -26 29 13 | 18.8 | 33600 | 110 | E,AC |
| 470 | 00 37 22.6 | -26 27 37 | 17.3 | 32896 | 68 | E |
| 470 | 00 37 24.3 | -26 28 23 | 20.0 | 32518 | 74 | E |
| 470 | 00 37 27.0 | -26 29 11 | 16.9 | 32968 | 29 | E |
| 470 | 00 37 27.4 | -26 30 14 | 19.5 | 32428 | 104 | E |
| 470 | 00 37 18.5 | -26 30 13 | 18.1 | 31987 | 26 | E |
| 470 | 00 37 22.1 | -26 30 32 | 19.9 | 34479 | 44 | E |
| 470 | 00 37 25.7 | -26 29 09 | 18.9 | 35105 | 38 | E |
| 471 | 00 37 49.8 | -24 57 43 | 20.3 | 33513 | 131 | E |
| 471 | 00 37 52.9 | -24 59 00 | 18.6 | 33264 | 86 | E |
| 471 | 00 37 54.9 | -24 59 27 | 18.2 | 33684 | 47 | E |
| 471 | 00 37 55.5 | -25 00 12 | 18.4 | 33261 | 53 | E |
| 471 | 00 37 56.5 | -24 58 34 | 19.2 | 33861 | 50 | E |
| 471 | 00 37 58.7 | -24 59 14 | 19.2 | 31433 | 74 | E |
| 471 | 00 38 01.2 | -25 00 17 | 18.2 | 34733 | 59 | E |
| 473 | 00 39 33.7 | -28 49 16 | 19.1 | 33112 | 50 | E |
| 473 | 00 39 31.6 | -28 47 34 | 18.6 | 32548 | 128 | E |
| 473 | 00 39 35.0 | -28 48 00 | 18.6 | 31322 | 35 | E |

| EDCC | RA | DEC | $b_j$ | cz | $\Delta$ cz | Notes |
|---|---|---|---|---|---|---|
| 473 | 00 39 37.2 | -28 48 26 | 18.1 | 31340 | 26 | E |
| 473 | 00 39 44.3 | -28 50 31 | 19.6 | 32218 | 44 | E |
| 473 | 00 39 41.5 | -28 48 35 | 16.1 | 32434 | 23 | E |
| 473 | 00 39 44.5 | -28 49 01 | 18.8 | 32818 | 44 | E |
| 473 | 00 39 46.4 | -28 49 26 | 19.0 | 31984 | 29 | E |
| 473 | 00 39 37.4 | -28 49 48 | 19.5 | 29274 | 95 | E |
| 473 | 00 39 44.3 | -28 47 35 | 18.0 | 33162 | 23 | E |
| 473 | 00 39 47.6 | -28 49 22 | 20.0 | 34194 | 56 | E |
| 473 | - | - | - | 32794 | 26 | E,NC |
| 473 | - | - | - | 32482 | 29 | E,NC |
| 473 | - | - | - | 33076 | 32 | E,NC |
| 474 | 00 40 20.5 | -26 20 06 | 18.7 | 34359 | 29 | E |
| 474 | 00 40 22.9 | -26 21 11 | 17.2 | 32896 | 23 | E |
| 474 | 00 40 25.0 | -26 21 02 | 19.3 | 33771 | 38 | E |
| 474 | 00 40 26.4 | -26 23 11 | 17.8 | 33807 | 38 | E |
| 474 | 00 40 27.8 | -26 21 27 | 17.5 | 33753 | 23 | E |
| 474 | 00 40 28.6 | -26 22 10 | 18.7 | 33774 | 56 | E |
| 474 | 00 40 24.5 | -26 20 58 | 19.6 | 33801 | 38 | E |
| 474 | 00 40 31.0 | -26 21 36 | 18.8 | 33795 | 68 | E |
| 474 | 00 40 32.1 | -26 21 53 | 18.8 | 30045 | 83 | E |
| 474 | 00 40 33.2 | -26 23 10 | 19.6 | 30272 | 32 | E |
| 474 | 00 41 09.6 | -26 14 01 | 17.8 | 16308 | 56 | E |
| 474 | 00 41 16.0 | -26 14 16 | 16.7 | 16362 | 35 | E |
| 474 | 00 41 17.7 | -26 17 05 | 17.4 | 30063 | 47 | E |
| 474 | 00 41 21.9 | -26 17 20 | 17.3 | 16224 | 35 | E |
| 474 | 00 41 28.7 | -26 17 42 | 19.3 | 36706 | 53 | E |
| 482 | 00 46 48.4 | -28 46 51 | 20.0 | 31022 | 53 | E,AC |
| 482 | 00 46 58.8 | -29 47 56 | 18.0 | 32278 | 29 | E |
| 482 | 00 46 58.8 | -29 47 56 | 18.0 | 32476 | 77 | E,AC |
| 482 | 00 46 58.1 | -29 47 33 | 17.2 | 31454 | 38 | E |
| 482 | 00 47 00.4 | -29 46 24 | 19.9 | 32839 | 23 | E |
| 482 | 00 47 00.2 | -29 46 24 | 18.0 | 31819 | 98 | E,AC |
| 482 | 00 47 03.4 | -29 48 05 | 18.8 | 33918 | 41 | E |
| 482 | 00 47 05.3 | -29 47 17 | 18.5 | 32812 | 29 | E,AC |
| 482 | 00 46 48.4 | -28 46 51 | 20.0 | 64910 | 29 | E,AC |
| 482 | 00 46 56.0 | -29 47 56 | 19.6 | 48641 | 44 | E |
| 482 | 00 46 54.4 | -29 44 56 | 18.2 | 32329 | 65 | A |
| 482 | 00 46 51.4 | -29 46 47 | 18.5 | 32353 | 47 | A |
| 482 | 00 46 59.2 | -29 47 44 | 18.0 | 32578 | 41 | A |
| 482 | 00 47 04.3 | -29 45 49 | 18.0 | 34068 | 32 | A |
| 482 | 00 47 17.0 | -29 50 36 | 17.9 | 32548 | 41 | A |
| 482 | 00 47 32.2 | -29 52 27 | 18.0 | 32290 | 29 | A |
| 482 | 00 47 03.0 | -29 50 29 | 18.3 | 32047 | 53 | A |
| 482 | 00 47 26.2 | -29 51 50 | 17.1 | 33162 | 83 | A |
| 482 | 00 46 53.9 | -29 49 24 | 18.7 | 32716 | 38 | A |
| 482 | 00 46 38.4 | -29 52 20 | 18.3 | 32476 | 47 | A |
| 482 | 00 46 29.5 | -29 57 22 | 18.1 | 32119 | 26 | A |
| 482 | 00 46 31.3 | -29 44 09 | 18.9 | 31897 | 44 | A |
| 482 | 00 46 40.1 | -29 42 52 | 18.6 | 31436 | 62 | A |
| 482 | 00 45 53.4 | -29 44 20 | 17.9 | 23018 | 89 | A |
| 482 | 00 45 59.5 | -29 45 11 | 18.3 | 22457 | 44 | A |
| 482 | 00 46 31.5 | -29 47 25 | 18.4 | 22511 | 59 | A |
| 482 | 00 46 33.3 | -29 53 47 | 17.4 | 22742 | 41 | A |
| 482 | 00 46 45.8 | -29 41 23 | 17.6 | 22820 | 32 | A |
| 482 | 00 46 56.5 | -29 51 15 | 18.4 | 22697 | 71 | A |
| 482 | 00 47 12.7 | -29 56 57 | 18.7 | 22706 | 26 | A |
| 485 | 00 48 49.7 | -28 47 51 | 17.5 | 33309 | 44 | E |
| 485 | 00 48 58.5 | -28 46 09 | 16.7 | 32956 | 50 | E |
| 485 | 00 49 00.8 | -28 46 35 | 18.1 | 33813 | 50 | E |
| 485 | 00 48 56.9 | -28 47 57 | 16.6 | 33981 | 26 | E |
| 485 | 00 48 44.6 | -28 46 16 | 17.2 | 34170 | 26 | E |
| 485 | 00 48 43.2 | -28 45 38 | 17.3 | 34152 | 74 | E |
| 494 | 00 53 11.8 | -37 32 10 | 17.2 | 17058 | 44 | E |
| 494 | 00 53 24.8 | -37 33 43 | 16.8 | 16455 | 35 | E |
| 494 | 00 53 24.0 | -37 35 22 | 16.5 | 16866 | 44 | E,AC |
| 494 | 00 53 24.1 | -37 40 44 | 15.4 | 16563 | 47 | E |
| 495 | 00 52 52.4 | -26 40 28 | 19.3 | 34173 | 23 | E |
| 495 | 00 53 01.4 | -26 39 27 | 18.9 | 33924 | 26 | E |
| 495 | 00 53 02.8 | -26 39 39 | 18.5 | 33900 | 35 | E |
| 495 | 00 53 04.0 | -26 40 00 | 18.1 | 33483 | 26 | E |
| 495 | 00 52 59.1 | -26 39 27 | 20.1 | 33816 | 32 | E |
| 495 | 00 53 06.4 | -26 40 17 | 19.0 | 33996 | 23 | E |
| 495 | 00 53 08.2 | -26 40 47 | 18.5 | 33696 | 53 | E |
| 495 | 00 53 19.6 | -26 36 19 | 17.4 | 35084 | 59 | E |
| 495 | 00 53 22.7 | -26 36 13 | 19.7 | 34395 | 68 | E |

| EDCC | RA | DEC | $b_j$ | cz | $\Delta$ cz | Notes |
|---|---|---|---|---|---|---|
| 495 | 00 53 17.6 | -26 38 38 | 19.6 | 34643 | 71 | E |
| 495 | 00 53 20.5 | -26 38 25 | 19.4 | 34712 | 44 | E |
| 495 | 00 53 24.1 | -26 38 09 | 17.3 | 34730 | 35 | E |
| 495 | 00 53 27.5 | -26 37 40 | 19.6 | 34275 | 35 | E |
| 495 | 00 53 28.3 | -26 37 52 | 18.3 | 33894 | 50 | E |
| 495 | 00 53 02.4 | -26 39 30 | 18.3 | 34494 | 104 | E |
| 495 | 00 53 28.6 | -26 38 00 | 18.9 | 33816 | 29 | E |
| 495 | 00 53 30.6 | -26 38 07 | 19.2 | 33720 | 74 | E |
| 499 | 00 53 39.4 | -38 10 26 | 16.6 | 35261 | 59 | E |
| 499 | 00 53 44.5 | -38 09 53 | 18.2 | 34973 | 44 | E |
| 499 | 00 53 26.5 | -38 12 41 | 18.1 | 34967 | 35 | E |
| 499 | 00 53 34.9 | -38 13 15 | 17.7 | 34302 | 44 | E |
| 500 | 00 54 03.4 | -30 19 37 | 18.6 | 33813 | 71 | E |
| 500 | 00 54 08.3 | -30 20 42 | 17.8 | 33984 | 53 | E |
| 500 | 00 54 13.6 | -30 19 38 | 18.2 | 34011 | 50 | E |
| 500 | 00 54 03.2 | -30 18 25 | 18.9 | 28675 | 56 | E |
| 500 | 00 54 05.9 | -30 22 02 | 18.2 | 47370 | 26 | E |
| 500 | 00 54 05.7 | -30 22 40 | 19.0 | 49705 | 50 | E |
| 500 | 00 54 09.8 | -30 18 47 | 20.5 | 64374 | 50 | E,AC |
| 519 | 01 02 34.7 | -40 01 45 | 17.8 | 31939 | 50 | A |
| 519 | 01 03 35.0 | -40 07 08 | 18.1 | 31616 | 62 | A |
| 519 | 01 03 02.5 | -40 10 56 | 17.7 | 31703 | 71 | A |
| 519 | 01 02 17.2 | -40 13 51 | 18.3 | 31951 | 50 | A |
| 519 | 01 02 07.5 | -40 10 48 | 18.1 | 32437 | 50 | A |
| 519 | 01 02 03.8 | -40 06 56 | 17.5 | 32761 | 110 | A |
| 519 | 01 01 45.7 | -40 03 04 | 16.6 | 31975 | 152 | A |
| 519 | 01 01 56.2 | -40 02 45 | 17.8 | 32659 | 59 | A |
| 519 | 01 01 50.1 | -40 01 20 | 18.4 | 31858 | 41 | A |
| 519 | 01 01 03.0 | -40 07 20 | 17.1 | 28279 | 83 | A |
| 519 | 01 01 21.1 | -40 08 46 | 16.5 | 9062 | 14 | A |
| 519 | 01 02 15.0 | -40 22 20 | 17.5 | 27838 | 41 | A |
| 519 | 01 02 52.2 | -40 10 45 | 17.9 | 27499 | 32 | A |
| 520 | 01 02 01.7 | -24 13 41 | 20.3 | 48302 | 59 | E |
| 520 | 01 02 02.0 | -24 13 26 | 19.3 | 49852 | 35 | E |
| 520 | 01 02 04.0 | -24 14 06 | 20.4 | 48908 | 53 | E |
| 520 | 01 02 04.4 | -24 13 37 | 19.2 | 47735 | 38 | E |
| 520 | 01 02 06.7 | -24 13 52 | 19.5 | 46935 | 50 | E,AC |
| 520 | 01 02 07.3 | -24 15 50 | 18.9 | 46929 | 29 | E |
| 520 | 01 02 08.3 | -24 14 32 | 17.3 | 47843 | 29 | E |
| 520 | 01 02 11.6 | -24 13 52 | 20.8 | 47960 | 89 | E |
| 520 | 01 02 06.6 | -24 15 53 | 19.2 | 56357 | 74 | E |
| 520 | 01 02 13.2 | -24 13 38 | 18.6 | 34539 | 104 | E |
| 520 | 01 02 14.3 | -24 13 56 | 20.6 | 57383 | 41 | E |
| 524 | 01 05 40.7 | -36 59 59 | 18.4 | 33960 | 147 | E |
| 524 | 01 05 45.0 | -37 03 23 | 19.6 | 35555 | 65 | E |
| 524 | 01 05 47.1 | -37 03 20 | 18.3 | 34266 | 41 | E |
| 524 | 01 05 48.0 | -37 00 36 | 18.1 | 34134 | 35 | E |
| 524 | 01 05 49.4 | -37 00 47 | 18.9 | 34098 | 38 | E |
| 524 | 01 05 41.7 | -37 02 04 | 18.9 | 36382 | 29 | E |
| 524 | 01 05 46.4 | -37 01 24 | 19.0 | 36241 | 56 | E |
| 524 | 01 05 50.2 | -37 02 20 | 19.6 | 35954 | 29 | E |
| 524 | 01 05 52.7 | -36 59 35 | 18.6 | 36487 | 35 | E |
| 526 | 01 06 30.5 | -40 37 54 | 19.2 | 42036 | 50 | E |
| 526 | 01 06 37.9 | -40 37 27 | 18.1 | 40912 | 47 | E |
| 526 | 01 06 24.0 | -40 35 53 | 19.0 | 41095 | 32 | E |
| 526 | 01 06 29.4 | -40 36 16 | 19.3 | 42522 | 56 | E |
| 526 | 01 06 21.5 | -40 36 14 | 18.0 | 43215 | 32 | E,AC |
| 526 | 01 06 21.5 | -40 35 10 | 18.8 | 43574 | 32 | E |
| 553 | 01 23 27.7 | -39 44 15 | 16.7 | 26399 | 23 | A |
| 553 | 01 24 17.8 | -39 45 09 | 18.5 | 26054 | 26 | A |
| 553 | 01 23 40.3 | -39 44 49 | 17.9 | 26051 | 35 | A |
| 553 | 01 24 36.8 | -39 50 08 | 18.6 | 26456 | 74 | A |
| 553 | 01 23 47.2 | -39 47 38 | 17.5 | 26843 | 26 | A |
| 553 | 01 23 48.5 | -39 48 58 | 18.2 | 26459 | 38 | A |
| 553 | 01 23 30.3 | -39 45 20 | 17.5 | 26366 | 29 | A |
| 553 | 01 23 21.9 | -39 44 43 | 18.5 | 26576 | 26 | A |
| 553 | 01 23 20.6 | -39 52 25 | 17.4 | 26567 | 47 | A |
| 553 | 01 23 07.6 | -39 45 20 | 17.7 | 24193 | 23 | A,AC |
| 553 | 01 23 16.7 | -39 45 31 | 17.9 | 26702 | 65 | A |
| 553 | 01 23 14.2 | -39 44 14 | 18.1 | 26645 | 80 | A |
| 553 | 01 23 10.4 | -39 44 22 | 17.9 | 26273 | 17 | A |
| 553 | 01 23 43.5 | -39 53 17 | 18.5 | 10504 | 26 | A |

| EDCC | RA | DEC | $b_j$ | cz | $\Delta$ cz | Notes |
|---|---|---|---|---|---|---|
| 555 | 01 23 48.2 | -29 49 51 | 18.3 | 42531 | 59 | E |
| 555 | 01 23 46.8 | -29 50 07 | 18.5 | 42474 | 68 | E |
| 555 | 01 23 45.6 | -29 50 23 | 18.4 | 42543 | 80 | E |
| 555 | 01 23 10.2 | -29 48 36 | 18.5 | 29040 | 41 | E |
| | | | | | | |
| 557 | 01 23 44.5 | -38 13 30 | 19.5 | 24067 | 299 | E |
| 557 | 01 23 49.3 | -38 11 58 | 18.8 | 23395 | 44 | E |
| 557 | 01 23 50.8 | -38 13 30 | 15.1 | 24094 | 35 | E,AC |
| 557 | 01 23 55.2 | -38 12 17 | 17.4 | 24307 | 38 | E |
| 557 | 01 23 56.7 | -38 14 40 | 18.1 | 23614 | 38 | E |
| 557 | 01 23 59.7 | -38 12 05 | 19.4 | 23872 | 29 | E |
| 557 | 01 23 47.9 | -38 13 03 | 18.9 | 25839 | 59 | E |
| 557 | 01 23 51.0 | -38 13 30 | 15.1 | 23095 | 17 | E,AC |
| | | | | | | |
| 570 | 01 30 14.9 | -42 22 06 | 16.9 | 26249 | 29 | E |
| 570 | 01 29 47.3 | -42 26 19 | 16.8 | 26250 | 23 | E |
| 570 | 01 29 49.2 | -42 25 47 | 17.9 | 26519 | 32 | E |
| 570 | 01 30 15.4 | -42 21 29 | 18.3 | 25074 | 44 | E |
| | | | | | | |
| 571 | 01 30 03.4 | -31 20 57 | 15.7 | 21336 | 47 | E |
| 571 | 01 30 11.6 | -31 19 05 | 17.1 | 21815 | 62 | E |
| 571 | 01 30 12.7 | -31 22 03 | 18.9 | 21719 | 83 | E |
| 571 | 01 30 05.9 | -31 20 24 | 17.2 | 20808 | 38 | E |
| 571 | 01 30 07.2 | -31 19 55 | 18.4 | 22847 | 98 | E |
| | | | | | | |
| 575 | 01 31 36.8 | -27 47 06 | 17.3 | 37021 | 35 | E |
| 575 | 01 31 34.8 | -27 46 48 | 18.2 | 37854 | 29 | E |
| 575 | 01 31 33.1 | -27 46 24 | 17.0 | 37357 | 35 | E |
| 575 | 01 31 45.0 | -27 46 29 | 18.0 | 36910 | 41 | E |
| 575 | 01 31 38.3 | -27 45 04 | 18.3 | 38904 | 29 | E |
| 575 | 01 31 30.4 | -27 44 49 | 18.1 | 29049 | 140 | E |
| 575 | 01 31 48.1 | -27 46 31 | 18.2 | 41086 | 41 | E |
| | | | | | | |
| 591 | 01 41 52.0 | -35 30 58 | 16.8 | 20895 | 26 | E |
| 591 | 01 41 55.0 | -35 34 01 | 17.1 | 19810 | 68 | E |
| 591 | 01 41 57.9 | -35 32 49 | 16.7 | 20442 | 32 | E |
| 591 | 01 41 54.3 | -35 34 03 | 17.8 | 19612 | 26 | E |
| 591 | 01 42 06.2 | -35 30 26 | 19.2 | 20598 | 80 | E |
| 591 | 01 41 57.2 | -35 33 50 | 19.6 | 70993 | 137 | E |
| 591 | 01 42 03.0 | -35 34 09 | 19.6 | 46176 | 41 | E |
| 591 | 01 42 05.2 | -35 30 51 | 20.1 | 70568 | 164 | E |

| EDCC | RA | DEC | $b_j$ | cz | $\Delta$ cz | Notes |
|---|---|---|---|---|---|---|
| 606 | 01 58 20.5 | -33 07 40 | 18.8 | 30815 | 41 | E |
| 606 | 01 58 20.3 | -33 07 20 | 19.1 | 31819 | 32 | E |
| 606 | 01 58 20.4 | -33 07 10 | 18.0 | 31217 | 38 | E |
| 606 | 01 58 20.2 | -33 06 52 | 18.4 | 31235 | 35 | E |
| 606 | 01 58 53.9 | -33 06 20 | 17.4 | 31493 | 47 | E |
| 606 | 01 58 51.2 | -33 06 17 | 18.8 | 31328 | 68 | E |
| 606 | 01 58 29.4 | -33 13 36 | 18.7 | 28510 | 23 | E |
| 606 | 01 58 33.4 | -33 15 05 | 18.0 | 29130 | 59 | E |
| 606 | 01 58 33.7 | -33 14 05 | 17.6 | 29028 | 26 | E |
| 606 | 01 58 35.2 | -33 14 32 | 17.4 | 28899 | 29 | E |
| 606 | 01 58 38.5 | -33 14 07 | 18.7 | 29136 | 32 | E |
| 606 | 01 58 49.1 | -33 06 18 | 19.5 | 18449 | 38 | E,AC |
| | | | | | | |
| 618 | 02 01 27.1 | -41 21 25 | 19.0 | 37096 | 56 | E,AC |
| 618 | 02 01 30.4 | -41 19 54 | 19.0 | 36157 | 50 | E |
| 618 | 02 01 31.4 | -41 20 25 | 16.9 | 38100 | 47 | E |
| 618 | 02 01 34.8 | -41 22 17 | 18.6 | 38412 | 35 | E |
| 618 | 02 01 37.6 | -41 21 00 | 20.0 | 37336 | 35 | E,AC |
| 618 | 02 01 38.6 | -41 21 41 | 20.3 | 37378 | 65 | E |
| 618 | 02 01 28.9 | -41 22 11 | 19.6 | 43574 | 68 | E |
| 618 | 02 01 33.1 | -41 20 07 | 18.0 | 35684 | 32 | E |
| 618 | 02 01 36.8 | -41 20 50 | 18.9 | 40268 | 44 | E |
| | | | | | | |
| 629 | 02 07 14.5 | -37 33 39 | 16.3 | 27126 | 53 | E |
| | | | | | | |
| 632 | 02 09 17.5 | -40 31 17 | 19.9 | 30389 | 134 | E |
| 632 | 02 09 23.7 | -40 31 42 | 18.9 | 31181 | 38 | E |
| 632 | 02 09 23.4 | -40 32 04 | 18.6 | 30350 | 38 | E |
| 632 | 02 09 18.5 | -40 29 50 | 20.0 | 31607 | 41 | E |
| 632 | 02 09 20.9 | -40 31 11 | 18.0 | 29538 | 41 | E |
| 632 | 02 09 22.2 | -40 32 07 | 19.0 | 29037 | 86 | E |
| 632 | 02 09 31.8 | -40 32 16 | 19.7 | 71650 | 37 | E |
| | | | | | | |
| 649 | 02 24 59.1 | -26 40 39 | 17.4 | 44030 | 44 | E |
| 649 | 02 25 01.8 | -26 40 06 | 18.3 | 44075 | 29 | E |
| 649 | 02 25 02.7 | -26 40 13 | 19.3 | 44219 | 83 | E |
| 649 | 02 25 03.5 | -26 41 34 | 18.7 | 43274 | 32 | E |
| 649 | 02 24 53.6 | -26 42 00 | 17.8 | 17789 | 23 | E |
| 649 | 02 25 03.4 | -26 41 05 | 17.4 | 24855 | 53 | E |
| 649 | 02 25 06.1 | -26 42 08 | 16.7 | 10222 | 44 | E |

| EDCC | RA | DEC | $b_j$ | cz | $\Delta$ cz | Notes |
|---|---|---|---|---|---|---|
| 653 | 02 28 42.4 | -33 19 31 | 16.4 | 23557 | 35 | E |
| 653 | 02 27 14.4 | -33 45 10 | 15.2 | 23131 | 35 | E,AC |
| 653 | 02 27 11.7 | -33 43 40 | 16.5 | 23356 | 29 | E,AC |
| 653 | 02 28 50.0 | -33 20 42 | 18.4 | 23995 | 59 | E |
| 653 | 02 28 05.6 | -33 33 23 | 17.6 | 24154 | 74 | E |
| 653 | 02 28 02.8 | -33 32 32 | 18.1 | 24343 | 101 | E |
| | | | | | | |
| 658 | 02 27 21.6 | -33 28 05 | 15.2 | 9635 | 194 | E |
| 658 | 02 29 57.4 | -33 11 41 | 18.4 | 24082 | 29 | A |
| 658 | 02 28 44.1 | -33 18 03 | 18.0 | 22004 | 59 | A |
| 658 | 02 30 01.5 | -33 19 06 | 18.1 | 21656 | 62 | A |
| 658 | 02 29 48.0 | -33 19 24 | 17.5 | 21267 | 32 | A |
| 658 | 02 29 40.6 | -33 19 50 | 18.1 | 23554 | 26 | A |
| 658 | 02 29 06.1 | -33 24 37 | 18.4 | 22667 | 20 | A |
| 658 | 02 28 40.3 | -33 24 13 | 18.5 | 22502 | 29 | A |
| 658 | 02 28 36.5 | -33 26 26 | 17.9 | 23413 | 50 | A |
| 658 | 02 28 02.8 | -33 32 32 | 18.1 | 23953 | 74 | A |
| 658 | 02 28 04.8 | -33 26 25 | 17.4 | 24214 | 65 | A |
| 658 | 02 28 04.6 | -33 13 50 | 18.2 | 11344 | 35 | A |
| 658 | 02 28 45.1 | -33 17 13 | 18.7 | 31274 | 29 | A |
| 658 | 02 28 48.1 | -33 24 57 | 17.9 | 33597 | 71 | A |
| 658 | 02 29 00.9 | -33 21 24 | 18.0 | 34071 | 71 | A |
| 658 | 02 28 42.4 | -33 19 31 | 16.4 | 23560 | 32 | A |
| 658 | 02 28 03.7 | -33 24 50 | 17.9 | 23317 | 17 | A |
| 658 | 02 27 48.8 | -33 23 57 | 16.3 | 23827 | 44 | A |
| 658 | 02 28 35.7 | -33 16 43 | 18.2 | 23317 | 26 | A |
| 658 | 02 28 08.9 | -33 12 31 | 17.4 | 23233 | 26 | A |
| 658 | 02 28 43.8 | -33 15 50 | 18.2 | 22325 | 17 | A |
| 658 | 02 28 29.3 | -33 11 27 | 18.5 | 22994 | 17 | A |
| 658 | 02 28 18.2 | -33 04 28 | 18.6 | 21638 | 29 | A |
| 658 | 02 28 28.0 | -33 04 00 | 17.5 | 21779 | 38 | A |
| 658 | 02 28 48.9 | -33 05 57 | 17.3 | 21923 | 71 | A |
| 658 | 02 29 12.0 | -33 11 29 | 18.1 | 21309 | 23 | A |
| | | | | | | |
| 683 | 02 42 57.9 | -26 37 39 | 18.5 | 40304 | 50 | E |
| 683 | 02 43 00.8 | -26 36 58 | 18.2 | 39875 | 83 | E |
| 683 | 02 42 09.7 | -26 26 50 | 18.2 | 40016 | 35 | E |
| 683 | 02 42 08.8 | -26 25 59 | 18.3 | 41290 | 47 | E |
| 693 | 02 45 49.3 | -42 03 59 | 17.8 | 21165 | 53 | E |
| 693 | 02 45 45.7 | -42 03 35 | 16.9 | 21357 | 65 | E |
| 693 | 02 45 38.3 | -42 02 40 | 17.4 | 41916 | 59 | E |
| | | | | | | |
| 699 | 02 49 02.7 | -25 08 07 | 19.8 | 33168 | 80 | E |
| 699 | 02 49 09.5 | -25 07 36 | 18.3 | 33112 | 53 | E |
| 699 | 02 49 22.1 | -25 15 02 | 17.8 | 33729 | 29 | E |
| 699 | 02 49 14.5 | -25 09 36 | 18.9 | 33252 | 41 | E |
| 699 | 02 49 05.1 | -25 07 50 | 19.3 | 10321 | 56 | E |
| 699 | 02 49 08.0 | -25 08 06 | 19.1 | 35435 | 83 | E |
| 699 | 02 49 16.5 | -25 10 37 | 19.6 | 34886 | 38 | E |
| | | | | | | |
| 710 | 02 53 44.6 | -22 52 00 | 17.7 | 37471 | 38 | E |
| 710 | 02 53 45.4 | -22 50 46 | 18.8 | 37539 | 62 | E |
| 710 | 02 53 38.4 | -22 54 00 | 19.0 | 37878 | 74 | E,AC |
| 710 | 02 53 41.7 | -22 52 48 | 18.8 | 36694 | 50 | E |
| | | | | | | |
| 712 | 02 54 25.5 | -24 54 36 | 16.3 | 33133 | 29 | A |
| 712 | 02 54 51.1 | -24 47 20 | 19.6 | 33396 | 53 | A |
| 712 | 02 54 36.4 | -24 52 48 | 17.5 | 32038 | 32 | A |
| 712 | 02 54 42.7 | -24 56 11 | 18.3 | 33396 | 50 | A |
| 712 | 02 54 34.7 | -24 55 10 | 19.3 | 33555 | 59 | A |
| 712 | 02 54 31.8 | -24 56 07 | 19.3 | 33399 | 32 | A |
| 712 | 02 54 21.3 | -24 58 23 | 18.3 | 32896 | 41 | A |
| 712 | 02 54 20.9 | -24 54 31 | 18.2 | 33324 | 146 | A,AC |
| 712 | 02 54 13.3 | -24 58 46 | 17.9 | 33876 | 80 | A |
| 712 | 02 54 18.4 | -24 53 45 | 19.2 | 34395 | 35 | A |
| 712 | 02 54 12.4 | -24 55 30 | 19.8 | 32518 | 95 | A |
| 712 | 02 54 19.7 | -24 52 57 | 19.5 | 33133 | 41 | A |
| 712 | 02 53 47.0 | -24 51 55 | 19.0 | 33270 | 50 | A |
| 712 | 02 54 26.6 | -24 49 19 | 18.9 | 47891 | 59 | A |
| 712 | 02 54 39.1 | -24 45 15 | 17.7 | 28798 | 44 | A |
| 712 | 02 54 43.1 | -24 52 06 | 19.8 | 37755 | 56 | A |
| 712 | 02 54 53.8 | -24 45 05 | 18.1 | 25815 | 32 | A |
| 712 | 02 54 54.1 | -24 56 55 | 17.4 | 38472 | 50 | A |
| 712 | 02 55 01.6 | -24 51 46 | 18.9 | 40340 | 38 | A |
| | | | | | | |
| 717 | 02 58 23.6 | -37 07 07 | 14.9 | 19681 | 41 | E |
| 717 | 02 58 30.3 | -37 14 51 | 15.6 | 20131 | 47 | E |
| 717 | 02 58 24.4 | -37 14 20 | 16.7 | 19687 | 62 | E |

| EDCC | RA | DEC | $b_j$ | cz | $\Delta$ cz | Notes |
|---|---|---|---|---|---|---|
| 722 | 03 00 44.6 | -37 04 42 | 18.6 | 20137 | 38 | E |
| 722 | 03 00 48.4 | -37 04 35 | 16.8 | 19819 | 32 | E |
| 722 | 03 00 39.6 | -37 06 47 | 17.6 | 19846 | 38 | E |
| 722 | 03 00 32.2 | -37 05 37 | 18.4 | 20116 | 44 | E |
| 726 | 03 04 45.8 | -38 56 38 | 17.5 | 26339 | 35 | E |
| 726 | 03 04 40.1 | -38 55 38 | 18.7 | 26066 | 50 | E |
| 726 | 03 05 01.5 | -39 03 32 | 19.1 | 26201 | 50 | E |
| 726 | 03 04 59.5 | -39 02 53 | 18.7 | 26165 | 20 | E |
| 728 | 03 06 40.6 | -36 56 12 | 16.2 | 19768 | 44 | A |
| 728 | 03 05 59.0 | -36 54 33 | 17.0 | 20131 | 38 | A |
| 728 | 03 05 33.7 | -36 58 35 | 17.2 | 19783 | 89 | A |
| 728 | 03 05 36.4 | -36 53 46 | 17.2 | 19972 | 35 | A |
| 728 | 03 04 41.6 | -36 49 55 | 17.2 | 19918 | 38 | A |
| 728 | 03 06 32.8 | -36 37 29 | 17.3 | 20526 | 98 | A |
| 728 | 03 06 17.7 | -36 46 31 | 18.3 | 20427 | 56 | A |
| 728 | 03 06 24.0 | -36 47 11 | 16.6 | 20673 | 65 | A |
| 728 | 03 05 13.6 | -36 36 25 | 16.2 | 13286 | 59 | A |
| 735 | 03 09 25.9 | -27 05 06 | 16.0 | 20526 | 44 | E |
| 735 | 03 09 09.8 | -27 05 13 | 19.3 | 20958 | 98 | E |
| 735 | 03 09 13.6 | -27 06 07 | 18.0 | 20403 | 17 | E |
| 735 | 03 09 16.4 | -27 07 10 | 16.0 | 20475 | 41 | E |
| 735 | 03 09 13.4 | -27 04 29 | 18.7 | 20403 | 47 | E |
| 735 | 03 09 20.6 | -27 06 25 | 18.9 | 20763 | 23 | E |
| 735 | 03 09 16.4 | -27 02 45 | 18.2 | 19717 | 41 | E |
| 735 | 03 09 15.1 | -27 04 57 | 17.6 | 34718 | 65 | E |
| 735 | 03 09 19.4 | -27 04 14 | 18.5 | 40429 | 77 | E |
| 742 | 03 11 49.0 | -38 29 11 | 15.2 | 25611 | 53 | E |
| 742 | 03 11 38.0 | -38 32 52 | 17.5 | 25443 | 68 | E,AC |
| 742 | 03 11 42.6 | -38 31 21 | 18.5 | 25623 | 47 | E |
| 742 | 03 11 51.0 | -38 28 39 | 18.9 | 25830 | 47 | E |
| 742 | 03 11 47.8 | -38 29 44 | 18.1 | 24064 | 101 | E |
| 742 | 03 11 41.1 | -38 32 54 | 17.2 | 24244 | 53 | E |
| 748 | 03 13 37.0 | -29 31 25 | 17.2 | 19897 | 71 | E |
| 748 | 03 13 30.3 | -29 30 37 | 16.7 | 19954 | 38 | E |
| 748 | 03 13 00.8 | -29 20 29 | 17.1 | 19891 | 35 | E |
| 748 | 03 12 53.4 | -29 21 10 | 17.7 | 20541 | 56 | E |
| 748 | 03 12 54.0 | -29 25 05 | 17.0 | 20176 | 47 | E,AC |
| 748 | 03 13 00.4 | -29 25 18 | 17.1 | 20208 | 53 | E |
| 758 | 03 20 31.9 | -41 32 25 | 15.7 | 19105 | 56 | E |
| 758 | 03 20 30.6 | -41 31 31 | 18.3 | 19387 | 62 | E |
| 758 | 03 20 35.5 | -41 31 41 | 18.2 | 18677 | 29 | E |
| 758 | 03 20 16.1 | -41 30 16 | 17.9 | 19144 | 56 | E |
| 758 | 03 20 25.5 | -41 33 39 | 20.2 | 19474 | 53 | E |
| 758 | 03 20 27.5 | -41 34 03 | 19.3 | 18662 | 62 | E |
| 758 | 03 20 28.5 | -41 31 11 | 15.5 | 19900 | 65 | E |
| 758 | 03 20 18.2 | -41 30 30 | 17.0 | 20673 | 29 | E |
| 758 | 03 20 30.5 | -41 34 09 | 19.7 | 20050 | 29 | E |
| 762 | 03 32 13.0 | -39 10 37 | 15.8 | 18680 | 29 | E |
| 762 | 03 32 16.5 | -39 09 33 | 16.0 | 18221 | 32 | E |
| 762 | 03 32 20.9 | -39 11 48 | 16.4 | 19009 | 35 | E |
| 762 | 03 32 20.5 | -39 09 21 | 17.9 | 20066 | 59 | E |
| 765 | 03 34 55.6 | -39 57 55 | 16.0 | 31247 | 38 | E |
| 765 | 03 35 20.0 | -39 58 50 | 18.6 | 31439 | 95 | E |
| 765 | 03 35 20.0 | -39 58 50 | 18.6 | 31834 | 56 | E |
| 765 | 03 35 20.4 | -39 58 30 | 17.2 | 31595 | 41 | E |
| 765 | 03 34 47.6 | -39 57 25 | 16.0 | 19954 | 35 | E,AC |
| 765 | 03 35 18.1 | -39 57 30 | 17.7 | 30255 | 38 | E |

Table 3: The galaxies observed in the EM cluster redshift survey at ESO and AAT. The columns represent: (1) The EDCC cluster identification name. (2) & (3) The equatorial coordinates (equinox 1950) from the EDSGC, RA and Dec respectively. (4) The approximate ($\pm 0.25\,\mathrm{mag}$) apparent magnitude of the galaxy in the photographic $b_j$ band from the EDSGC. (5) The heliocentric velocity of each galaxy in units of $\mathrm{km\,s^{-1}}$. (6) The internal redshift error in units of $\mathrm{km\,s^{-1}}$ from the cross–correlation routine, described in section 4.2. (7) An "E" in this column denotes an ESO observation, "A" denotes an AAT observation, and an "AC" denotes a larger than normal error in the coordinates and magnitude. This is the case for galaxies which are not detected in the EDSGC, like e.g. objects merged with other object images in the digitization process. 2 galaxies in E178, 3 in E261, 2 in E424 and 3 in E473 do not have a certain identification in our logfiles (indicated as "NC"). We prefer therefore not to list any coordinate and magnitude which would be too approximate, although we give the redshift and the information that the object belongs to that cluster.

Table 4

| EDCC | RA | DEC | $N_{tot}$ | $N_{clus}$ | cz | $\Delta$ cz | $\sigma_v$ | Notes |
|---|---|---|---|---|---|---|---|---|
| 5 | 21 29 33.9 | -22 53 02 | 6 | 3 | 33363 | 338 | - | |
| 42 | 21 46 21.9 | -30 56 38 | 8 | 4 | 35294 | 103 | - | |
| 51 | 21 49 22.2 | -29 08 02 | 6 | 5 | 27689 | 163 | - | |
| 57 | 21 53 30.9 | -30 23 00 | 4 | 4 | 27754 | 97 | - | |
| 80 | 21 59 25.2 | -22 48 38 | 36 | 29 | 21022 | 153 | 768(+128,-86) | |
| 99 | 22 06 35.0 | -27 33 24 | 4 | - | - | - | - | p |
| 114 | 22 11 08.7 | -36 54 40 | 2 | 2 | 10183 | - | - | |
| 115 | 22 11 09.1 | -35 13 31 | 4 | 3 | 21884 | 56 | - | |
| 124 | 22 14 43.9 | -35 57 33 | 10 | 9 | 44188 | 388 | 1011(+397,-183) | |
| 127 | 22 15 41.0 | -39 08 56 | 14 | - | - | - | - | p |
| 131 | 22 16 39.2 | -34 56 27 | 4 | 4 | 46884 | 266 | - | |
| 145 | 22 24 56.7 | -30 51 11 | 12 | 11 | 17050 | 353 | 1105(+367,-184) | |
| 172 | 22 35 50.4 | -37 09 50 | 4 | 3 | 17542 | 284 | - | |
| 175 | 22 36 27.4 | -38 05 29 | 4 | 3 | 46031 | 180 | - | |
| 178 | 22 37 50.6 | -34 14 33 | 2 | 2 | 14905 | 56 | - | |
| 198 | 22 46 29.7 | -41 10 02 | 6 | - | - | - | - | p |
| 201 | 22 47 13.5 | -31 26 26 | 4 | - | - | - | - | p |
| 216 | 22 50 40.5 | -25 47 14 | 5 | - | - | - | - | p |
| 230 | 22 56 32.8 | -31 07 12 | 4 | 2 | 32916 | 611 | - | |
| 235 | 22 59 55.3 | -33 38 06 | 4 | - | - | - | - | p |
| 247 | 23 02 56.1 | -39 22 15 | 4 | 3 | 50116 | 335 | - | |
| 256 | 23 07 36.2 | -23 12 40 | 4 | - | - | - | - | p |
| 261 | 23 09 09.0 | -29 19 41 | 8 | 6 | 35614 | 375 | 820(+477,-173) | |
| 269 | 23 12 32.5 | -38 02 09 | 12 | - | - | - | - | p |
| 285 | 23 18 12.7 | -42 09 25 | 4 | - | - | - | - | p |
| 297 | 23 23 46.2 | -24 14 59 | 5 | 3 | 26752 | 186 | - | |
| 307 | 23 27 31.1 | -39 33 25 | 3 | 3 | 16368 | 309 | - | |
| 311 | 23 28 36.0 | -36 47 41 | 7 | 5 | 28829 | 217 | - | |
| 316 | 23 29 49.8 | -36 31 28 | 4 | 4 | 28359 | 163 | - | |
| 326 | 23 35 09.7 | -38 29 19 | 7 | 3 | 32592 | 110 | - | |
| 332 | 23 39 04.3 | -29 28 46 | 3 | 3 | 15517 | 150 | - | |
| 348 | 23 44 33.7 | -28 31 41 | 33 | 32 | 8757 | 151 | 830(+132,-90) | |
| 366 | 23 52 19.6 | -27 56 40 | 8 | 6 | 22040 | 204 | 464(+271,-100) | |
| 372 | 23 54 18.2 | -34 54 40 | 1 | 1 | 12675 | - | - | |
| 392 | 00 00 13.9 | -34 56 38 | 8 | 7 | 33536 | 221 | 525(+261,-105) | |
| 394 | 00 00 32.1 | -36 12 59 | 6 | 6 | 14719 | 257 | 598(+348,-127) | |
| 400 | 00 03 39.1 | -34 58 49 | 10 | 7 | 34436 | 516 | 1222(+606,-243) | |
| 408 | 00 07 27.8 | -35 56 08 | 14 | 13 | 35864 | 272 | 871(+252,-136) | |

| EDCC | RA | DEC | $N_{tot}$ | $N_{clus}$ | cz | $\Delta$ cz | $\sigma_v$ | Notes |
|---|---|---|---|---|---|---|---|---|
| 410 | 00 08 46.0 | -29 07 35 | 2 | 2 | 18444 | - | - | |
| 418 | 00 12 30.2 | -24 09 59 | 9 | 8 | 19714 | 267 | 706(+309,-134) | |
| 419 | 00 12 49.7 | -26 21 10 | 5 | 3 | 37215 | 955 | - | |
| 421 | 00 13 35.9 | -35 13 53 | 7 | 5 | 43825 | 601 | - | |
| 424 | 00 14 05.9 | -34 07 59 | 2 | 2 | 22629 | - | - | |
| 429 | 00 15 23.2 | -35 25 03 | 19 | 17 | 29060 | 211 | 790(+188,-110) | |
| 437 | 00 18 01.2 | -25 54 26 | 6 | 5 | 18997 | 426 | - | |
| 438 | 00 20 23.5 | -38 24 13 | 4 | 4 | 35797 | 401 | - | |
| 447 | 00 26 07.8 | -23 54 04 | 6 | 4 | 33849 | 275 | - | |
| 448 | 00 26 34.0 | -30 26 27 | 17 | - | - | - | - | p |
| 450 | 00 27 23.4 | -29 45 01 | 5 | 4 | 29635 | 482 | - | |
| 460 | 00 34 47.1 | -28 44 47 | 31 | 21 | 33610 | 170 | 700(+145,-90) | |
| 462 | 00 35 14.5 | -39 23 42 | 12 | 11 | 18864 | 183 | 569(+186,-95) | |
| 470 | 00 37 26.4 | -26 26 25 | 10 | 7 | 32903 | 175 | 415(+206,-83) | |
| 471 | 00 37 43.3 | -24 56 48 | 7 | 4 | 33519 | 262 | - | |
| 473 | 00 40 03.7 | -28 50 23 | 14 | 13 | 32577 | 214 | 695(+200,-108) | |
| 474 | 00 40 44.7 | -26 19 54 | 15 | 8 | 33745 | 141 | 354(+156,-69) | |
| 482 | 00 46 50.3 | -29 47 22 | 30 | 21 | 32412 | 159 | 675(+138,-86) | |
| 485 | 00 48 56.3 | -28 46 50 | 6 | 6 | 33731 | 202 | 443(+258,-94) | |
| 494 | 00 53 24.9 | -37 36 36 | 4 | 4 | 16736 | 138 | - | |
| 495 | 00 53 28.6 | -26 36 09 | 17 | 14 | 34213 | 121 | 403(+110,-61) | |
| 499 | 00 53 51.4 | -38 10 02 | 4 | 4 | 34877 | 203 | - | |
| 500 | 00 53 59.7 | -30 20 02 | 7 | 3 | 33937 | 62 | - | |
| 519 | 01 02 07.7 | -40 06 22 | 13 | 9 | 32100 | 138 | 371(+147,-69) | |
| 520 | 01 02 11.5 | -24 15 37 | 11 | 8 | 48025 | 370 | 902(+394,-170) | |
| 524 | 01 05 39.9 | -37 01 21 | 9 | 9 | 35232 | 364 | 975(+383,-176) | |
| 526 | 01 06 30.7 | -40 37 20 | 6 | 6 | 42224 | 443 | 950(+552,-199) | |
| 553 | 01 23 09.0 | -39 41 37 | 14 | 12 | 26449 | 70 | 242(+67,-39) | |
| 555 | 01 23 24.9 | -29 48 28 | 4 | 3 | 42518 | 22 | - | |
| 557 | 01 23 46.9 | -38 14 35 | 8 | 6 | 23729 | 183 | 413(+240,-87) | |
| 570 | 01 30 07.7 | -42 24 39 | 4 | 4 | 26024 | 323 | - | |
| 571 | 01 30 15.4 | -31 20 04 | 5 | 4 | 21421 | 230 | - | |
| 575 | 01 31 52.4 | -27 47 19 | 7 | 5 | 37610 | 362 | - | |
| 591 | 01 41 46.0 | -35 32 50 | 8 | 5 | 20272 | 242 | - | |
| 606 | 01 58 27.2 | -33 11 15 | 12 | 11 | 30237 | 383 | 1152(+377,-190) | |
| 618 | 02 01 28.8 | -41 20 45 | 9 | 7 | 37166 | 368 | 865(+429,-172) | |
| 629 | 02 07 12.0 | -37 36 41 | 1 | 1 | 27126 | - | - | |
| 632 | 02 09 33.9 | -40 31 42 | 7 | 6 | 30351 | 393 | 872(+507,-184) | |

| EDCC | RA | DEC | $N_{tot}$ | $N_{clus}$ | cz | $\Delta$ cz | $\sigma_v$ | Notes |
|---|---|---|---|---|---|---|---|---|
| 649 | 02 24 54.3 | -26 43 07 | 7 | 4 | 43900 | 212 | - | |
| 653 | 02 27 16.9 | -33 41 37 | 6 | 6 | 23756 | 195 | 440(+257,-94) | |
| 658 | 02 28 34.9 | -33 17 56 | 26 | 22 | 22893 | 224 | 977(+194,-122) | |
| 683 | 02 42 26.6 | -26 31 11 | 4 | 4 | 40373 | 319 | - | |
| 693 | 02 45 51.1 | -42 03 38 | 3 | 2 | 21261 | - | - | |
| 699 | 02 49 17.9 | -25 09 01 | 7 | 6 | 33931 | 404 | 891(+518,-188) | |
| 710 | 02 53 42.0 | -22 51 30 | 4 | 4 | 37397 | 250 | - | |
| 712 | 02 54 19.6 | -24 55 51 | 19 | 13 | 33256 | 161 | 519(+152,-81) | |
| 717 | 02 58 55.5 | -37 14 57 | 3 | 3 | 19834 | 149 | - | |
| 722 | 03 01 03.9 | -37 07 47 | 4 | 4 | 19980 | 85 | | |
| 726 | 03 04 43.0 | -39 01 47 | 4 | 4 | 26194 | 56 | - | |
| 728 | 03 06 13.3 | -36 53 32 | 9 | 8 | 20151 | 124 | 323(+142,-64) | |
| 735 | 03 09 23.7 | -27 05 34 | 9 | 7 | 20464 | 146 | 359(+179,-73) | |
| 742 | 03 11 52.2 | -38 30 35 | 6 | 6 | 25137 | 315 | 709(+413,-151) | |
| 748 | 03 13 09.5 | -29 24 14 | 6 | 5 | 20026 | 69 | - | |
| 758 | 03 20 22.1 | -41 30 47 | 9 | 9 | 19453 | 220 | 618(+243,-112) | |
| 762 | 03 32 15.2 | -39 08 09 | 4 | 4 | 18994 | 392 | - | |
| 765 | 03 34 50.6 | -39 53 16 | 6 | 4 | 31530 | 124 | - | |

Table 4: The clusters observed in the EM redshift survey. The columns represent: (1) The EDCC cluster identification name. (2) & (3) The equatorial RA and Dec coordinates (equinox 1950) of the cluster centroid, taken from Lumsden *et al.* 1992. (4) The total number of redshifts taken for each cluster (listed in Table 3). (5) The number of cluster members determined from the de–projection algorithm described in section 5.1. (6) The mean heliocentric redshift of the galaxy members in each clusters in units of $km\,s^{-1}$. (7) The $1\sigma$ errors on the mean redshift of each cluster in units of $km\,s^{-1}$. (8) The radial velocity dispersions for clusters with $N_{clus} \geq 6$ in units of $km\,s^{-1}$, along with the $1\sigma$ errors (see section 6.1). (9) A "p" in this column indicates that the cluster has been classified as a projection effect (see section 5.2).

Table 5

| EDCC | RA | DEC | $R_1$ | $m_{10}$ | $\theta_A$ | PA | $\epsilon$ | cz | N | Reference |
|---|---|---|---|---|---|---|---|---|---|---|
| 57 | 21 53 30.9 | -30 23 00 | 59 | 18.1 | 0.172 | 32 | 0.24 | 27754 | 4 | EM |
| 72 | 21 56 56.9 | -42 20 24 | 30 | 18.3 | 0.157 | 83 | 0.57 | - | - | |
| 80 | 21 59 25.2 | -22 48 39 | 26 | 17.7 | 0.202 | - | - | 21022 | 29 | EM |
| 86 | 22 02 26.5 | -30 45 48 | 26 | 18.7 | 0.136 | 138 | 0.45 | 28180 | 1 | Loveday (1991) |
| 114 | 22 11 08.7 | -36 54 40 | 24 | 17.2 | 0.245 | 65 | 0.57 | 10183 | 2 | EM |
| 124 | 22 14 43.9 | -35 57 33 | 75 | 18.7 | 0.127 | 115 | 0.60 | 44188 | 9 | EM |
| 127 | 22 15 41.0 | -39 08 56 | 67 | 18.5 | 0.145 | - | - | - | - | EM (projection) |
| 129 | 22 16 02.1 | -24 26 56 | 28 | 16.8 | 0.30 | 97 | 0.18 | 10463 | 2 | Loveday (1991) |
| 145 | 22 24 56.7 | -30 51 11 | 35 | 17. | 0.271 | 132 | 0.21 | 17050 | 11 | EM |
| 160 | 22 31 32.4 | -37 59 54 | 77 | 18.7 | 0.138 | 11 | 0.22 | 45269 | 70 | Teague et al. (1990) |
| 165 | 22 33 35.6 | -24 36 25 | 25 | 17.7 | 0.197 | 116 | 0.40 | - | - | |
| 172 | 22 35 50.4 | -37 09 50 | 25 | 17.3 | 0.357 | 89 | 0.73 | 17542 | 3 | EM |
| 173 | 22 35 57.3 | -36 39 56 | 31 | 18.0 | 0.265 | 77 | 0.74 | 17058 | | Peacock (priv. comm.) |
| 175 | 22 36 27.4 | -38 05 29 | 34 | 18.5 | 0.148 | 179 | 0.77 | 46031 | 3 | EM |
| 176 | 22 36 54.3 | -36 26 25 | 35 | 18.3 | 0.237 | - | - | 17088 | 1 | Nicholson (1991) |
| 178 | 22 37 50.6 | -34 14 33 | 28 | 18.2 | 0.161 | 24 | 0.61 | 14905 | 2 | EM |
| 188 | 22 43 39.4 | -36 21 47 | 35 | 17.5 | 0.219 | 139 | 0.49 | 20116 | 3 | Muriel et al. 1990 |
| 201 | 22 47 13.5 | -31 26 26 | 37 | 18.4 | 0.155 | - | - | - | - | EM (projection) |
| 203 | 22 48 13.2 | -26 20 35 | 34 | 18.3 | 0.153 | 168 | 0.51 | - | - | |
| 216 | 22 50 40.5 | -25 47 14 | 58 | 18.2 | 0.163 | - | - | - | - | EM (projection) |
| 230 | 22 56 32.8 | -31 07 12 | 33 | 18.6 | 0.141 | - | - | 32916 | 2 | EM |
| 258 | 23 08 01.3 | -22 55 15 | 35 | 18.7 | 0.131 | 41 | 0.39 | 33547 | 1 | Ciardullo et al. 1985 |
| 260 | 23 08 28.2 | -28 59 43 | 30 | 18.7 | 0.131 | 103 | 0.66 | - | - | |
| 261 | 23 09 09.4 | -29 19 41 | 22 | 18.4 | 0.152 | 147 | 0.37 | 35614 | 6 | EM |
| 263 | 23 09 23.8 | -41 02 21 | 29 | 18.7 | 0.133 | 86 | 0.31 | - | - | |
| 269 | 23 12 32.5 | -38 02 09 | 44 | 18.6 | 0.140 | - | - | - | - | EM (projection) |
| 279 | 23 15 47.7 | -27 47 17 | 22 | 18.6 | 0.138 | 177 | 0.45 | 25182 | 1 | Nicholson (1991) |
| 285 | 23 18 12.7 | -42 09 25 | 30 | 17.7 | 0.199 | - | - | - | - | EM(projection) |
| 297 | 23 23 46.2 | -24 14 59 | 42 | 18.3 | 0.159 | 68 | 0.76 | 26752 | 3 | EM |
| 311 | 23 28 36.0 | -36 47 41 | 43 | 18.1 | 0.253 | - | - | 28829 | 5 | EM |
| 316 | 23 29 49.8 | -36 31 28 | 49 | 17.4 | 0.343 | 43 | 0.59 | 28359 | 4 | EM |
| 332 | 23 39 04.3 | -29 28 46 | 34 | 17.8 | 0.194 | 54 | 0.30 | 15517 | 3 | EM |
| 335 | 23 39 36.3 | -30 28 45 | 23 | 18.7 | 0.136 | 102 | 0.61 | 21015 | 1 | Olowin et al. (1988) |
| 348 | 23 44 33.7 | -28 31 41 | 22 | 16.2 | 0.430 | 130 | 0.44 | 8757 | 32 | EM |
| 366 | 23 52 19.6 | -27 56 40 | 33 | 17.7 | 0.205 | 153 | 0.32 | 22040 | 6 | EM |
| 377 | 23 56 26.4 | -32 09 32 | 53 | 17.8 | 0.185 | 29 | 0.41 | 17748 | 1 | Broadbent 1992 (priv. comm.) |
| 381 | 23 57 14.2 | -39 45 34 | 22 | 18.5 | 0.145 | 11 | 0.52 | 30455 | 1 | Muriel et al. (1991) |
| 387 | 23 58 35.2 | -36 39 00 | 27 | 18.6 | 0.134 | 73 | 0.28 | - | - | |
| 392 | 00 00 13.9 | -34 56 38 | 34 | 18.6 | 0.140 | 128 | 0.28 | 33536 | 7 | EM |
| 394 | 00 00 32.1 | -36 12 59 | 24 | 17.4 | 0.227 | 148 | 0.21 | 14719 | 6 | EM |

| EDCC | RA | DEC | $R_1$ | $m_{10}$ | $\theta_A$ | PA | $\epsilon$ | cz | N | Reference |
|---|---|---|---|---|---|---|---|---|---|---|
| 396 | 00 00 37.9 | -27 28 16 | 22 | 18.3 | 0.159 | 116 | 0.61 | 20416 | 3 | Postman et al. (1992) |
| 398 | 00 01 25.3 | -23 25 16 | 29 | 18.6 | 0.140 | 98 | 0.33 | - | - | |
| 400 | 00 03 39.1 | -34 58 49 | 80 | 18.5 | 0.147 | 170 | 0.21 | 34436 | 7 | EM |
| 407 | 00 06 38.9 | -35 35 00 | 25 | 17.7 | 0.196 | 144 | 0.21 | 14870 | 3 | Huchra et al. (1992) |
| 410 | 00 08 46.0 | -29 07 35 | 25 | 17.4 | 0.225 | 172 | 0.23 | 18444 | 2 | EM |
| 418 | 00 12 30.2 | -24 09 59 | 32 | 17.1 | 0.264 | 35 | 0.22 | 19714 | 8 | EM |
| 419 | 00 12 49.7 | -26 21 10 | 35 | 18.7 | 0.133 | 74 | 0.70 | 37215 | 3 | EM |
| 424 | 00 14 05.9 | -34 07 59 | 22 | 18.4 | 0.141 | 11 | 0.23 | 22629 | 2 | EM |
| 429 | 00 15 23.2 | -35 25 03 | 63 | 18.0 | 0.176 | 4 | 0.56 | 29060 | 17 | EM |
| 431 | 00 16 17.4 | -42 01 24 | 30 | 18.2 | 0.166 | 79 | 0.73 | 27671 | 3 | Muriel et al. (1991) |
| 437 | 00 18 01.2 | -25 54 26 | 59 | 18.5 | 0.145 | 95 | 0.57 | 18998 | 5 | EM |
| 446 | 00 25 44.0 | -35 43 05 | 22 | 18.7 | 0.130 | 125 | 0.81 | - | - | |
| 447 | 00 26 07.8 | -23 54 04 | 46 | 18.4 | 0.152 | 3 | 0.49 | 33849 | 4 | EM |
| 448 | 00 26 34.0 | -30 26 27 | 22 | 18.7 | 0.181 | - | - | - | - | EM (projection) |
| 450 | 00 27 23.4 | -29 45 01 | 26 | 18.5 | 0.143 | 54 | 0.38 | 29635 | 4 | EM |
| 460 | 00 34 47.1 | -28 44 47 | 30 | 17.9 | 0.183 | 62 | 0.10 | 33610 | 21 | EM |
| 462 | 00 35 14.5 | -39 23 42 | 35 | 18.5 | 0.147 | 49 | 0.62 | 18864 | 11 | EM |
| 473 | 00 40 03.7 | -28 50 23 | 44 | 18.0 | 0.265 | 112 | 0.62 | 32577 | 13 | EM |
| 474 | 00 40 44.7 | -26 19 54 | 22 | 18.6 | 0.203 | 7 | 0.43 | 33745 | 8 | EM |
| 480 | 00 45 56.6 | -42 17 21 | 23 | 18.5 | 0.141 | 40 | 0.20 | - | 1 | Huchra (priv. comm.) |
| 482 | 00 46 50.3 | -29 47 22 | 54 | 18.0 | 0.177 | 139 | 0.24 | 32412 | 21 | EM |
| 495 | 00 53 28.6 | -26 36 09 | 55 | 18.1 | 0.176 | 62 | 0.16 | 34213 | 14 | EM |
| 500 | 00 53 59.7 | -30 20 02 | 40 | 18.8 | 0.134 | 37 | 0.62 | 33937 | 3 | EM |
| 505 | 00 54 38.9 | -29 57 01 | 26 | 18.5 | 0.147 | - | - | - | - | |
| 507 | 00 54 55.5 | -26 33 36 | 23 | 18.4 | 0.155 | 143 | 0.26 | 33811 | 1 | Peterson et al. (1986) |
| 519 | 01 02 07.7 | -40 06 22 | 37 | 18.2 | 0.164 | 8 | 0.43 | 32100 | 9 | EM |
| 520 | 01 02 11.5 | -24 15 37 | 70 | 18.6 | 0.138 | 125 | 0.19 | 48025 | 8 | EM |
| 524 | 01 05 39.9 | -37 01 21 | 55 | 18.6 | 0.137 | 153 | 0.34 | 35232 | 9 | EM |
| 546 | 01 19 45.7 | -39 53 30 | 27 | 18.6 | 0.141 | 157 | 0.54 | - | - | |
| 553 | 01 23 09.0 | -39 41 37 | 25 | 18.1 | 0.168 | 146 | 0.53 | 26449 | 12 | EM |
| 555 | 01 23 24.9 | -29 48 28 | 22 | 18.4 | 0.148 | 161 | 0.59 | 42518 | 3 | EM |
| 557 | 01 23 46.9 | -38 14 35 | 42 | 18.1 | 0.168 | 39 | 0.29 | 23729 | 6 | EM |
| 570 | 01 30 07.7 | -42 24 39 | 29 | 17.9 | 0.185 | 31 | 0.46 | 26024 | 4 | EM |
| 575 | 01 31 52.4 | -27 47 19 | 29 | 18.0 | 0.175 | 44 | 0.27 | 37610 | 5 | EM |
| 587 | 01 39 46.9 | -42 24 58 | 31 | 17.6 | 0.214 | 18 | 0.42 | 22790 | | Dalton et al 1994 |
| 591 | 01 41 46.0 | -35 32 50 | 28 | 17.6 | 0.211 | 2 | 0.22 | 20272 | 5 | EM |
| 606 | 01 58 27.2 | -33 11 15 | 33 | 18. | 0.179 | 62 | 0.43 | 30237 | 11 | EM |
| 618 | 02 01 28.8 | -41 20 45 | 30 | 18.4 | 0.153 | 122 | 0.41 | 37166 | 7 | EM |
| 645 | 02 23 11.7 | -29 43 25 | 30 | 18.1 | 0.170 | 161 | 0.56 | - | - | |
| 653 | 02 27 16.9 | -33 41 37 | 30 | 18.3 | 0.153 | 67 | 0.33 | 23756 | 6 | EM |

| EDCC | RA | DEC | $R_1$ | $m_{10}$ | $\theta_A$ | PA | $\epsilon$ | cz | N | Reference |
|---|---|---|---|---|---|---|---|---|---|---|
| 658 | 02 28 34.9 | -33 17 56 | 43 | 18.5 | 0.141 | 79 | 0.48 | 22893 | 22 | EM |
| 683 | 02 42 26.6 | -26 31 11 | 22 | 18.3 | 0.162 | 136 | 0.59 | 40373 | 4 | EM |
| 699 | 02 49 17.9 | -25 09 01 | 65 | 18.2 | 0.163 | 119 | 0.10 | 33931 | 6 | EM |
| 700 | 02 49 39.5 | -25 47 16 | 27 | 18.4 | 0.151 | 121 | 0.56 | 33577 | | Peacock (priv. comm.) |
| 707 | 02 51 30.2 | -33 41 40 | 26 | 17.9 | 0.187 | 8 | 0.62 | - | - | |
| 712 | 02 54 19.6 | -24 55 51 | 26 | 18.6 | 0.138 | 19 | 0.53 | 33256 | 13 | EM |
| 717 | 02 58 55.5 | -37 14 57 | 43 | 17.3 | 0.241 | 144 | 0.63 | 19834 | 3 | EM |
| 726 | 03 04 43.0 | -39 01 47 | 26 | 18.5 | 0.143 | 158 | 0.44 | 26194 | 4 | EM |
| 728 | 03 06 13.3 | -36 53 32 | 38 | 17.3 | 0.237 | 34 | 0.16 | 20151 | 8 | EM |
| 729 | 03 06 15.5 | -23 53 09 | 24 | 17.4 | 0.225 | 107 | 0.13 | 19736 | 2 | Postman *et al.* (1992) |
| 735 | 03 09 23.7 | -27 05 34 | 22 | 17.4 | 0.230 | 146 | 0.65 | 20464 | 7 | EM |
| 742 | 03 11 52.2 | -38 30 35 | 25 | 17.9 | 0.187 | 172 | 0.39 | 25137 | 6 | EM |
| 748 | 03 13 09.5 | -29 24 14 | 29 | 17.7 | 0.202 | 106 | 0.49 | 20026 | 5 | EM |
| 758 | 03 20 22.1 | -41 30 47 | 36 | 17.9 | 0.187 | 41 | 0.13 | 19453 | 9 | EM |
| 762 | 03 32 15.2 | -39 08 09 | 23 | 17.3 | 0.236 | 68 | 0.31 | 18994 | 4 | EM |
| 763 | 03 32 55.2 | -39 38 01 | 25 | 18.2 | 0.152 | 88 | 0.63 | 30879 | | Peacock (priv. comm.) |
| 765 | 03 34 50.6 | -39 53 16 | 22 | 18.0 | 0.177 | 134 | 0.18 | 31530 | 4 | EM |

Table 5: The list of clusters selected from the EDSGC using a reduced Abell radius of $1.0\,h^{-1}$Mpc and subject to the selection criteria given in section 7, used as the basis for the clustering analysis of Nichol *et al.* 1992 (see section 7) and Martin *et al.* 1994. The columns represent: (1)–(3) The EDCC identification and coordinates as given in Table 4. (4) The number of galaxy members between the magnitudes $m_3$ and $m_3+2$ calculated using an Abell radius of $1\,h^{-1}$Mpc. (5) Magnitude of the $10^{th}$ brightest galaxy, after background correction, within the $1\,h^{-1}$Mpc radius. (6) The angular 'reduced Abell radius' calculated using the equivalent of Abell's $m_{10}$ distance estimate. (7) The position angle of the cluster, in degrees measured from the North clockwise, as defined in Martin *et al.* (1994). This parameter is not used in this paper but is included here for completeness. (8) The eccentricity of the clusters defined as the ratio between the minor and the major axis of the cluster, as defined and used in Martin *et al.* 1994. (9) The radial velocity of the cluster in units of $\text{km}\,\text{s}^{-1}$. (8) The number of galaxies used to determine cz. For the EM clusters, this is just $N_{\text{clus}}$. (10) The source of the cluster radial velocity.

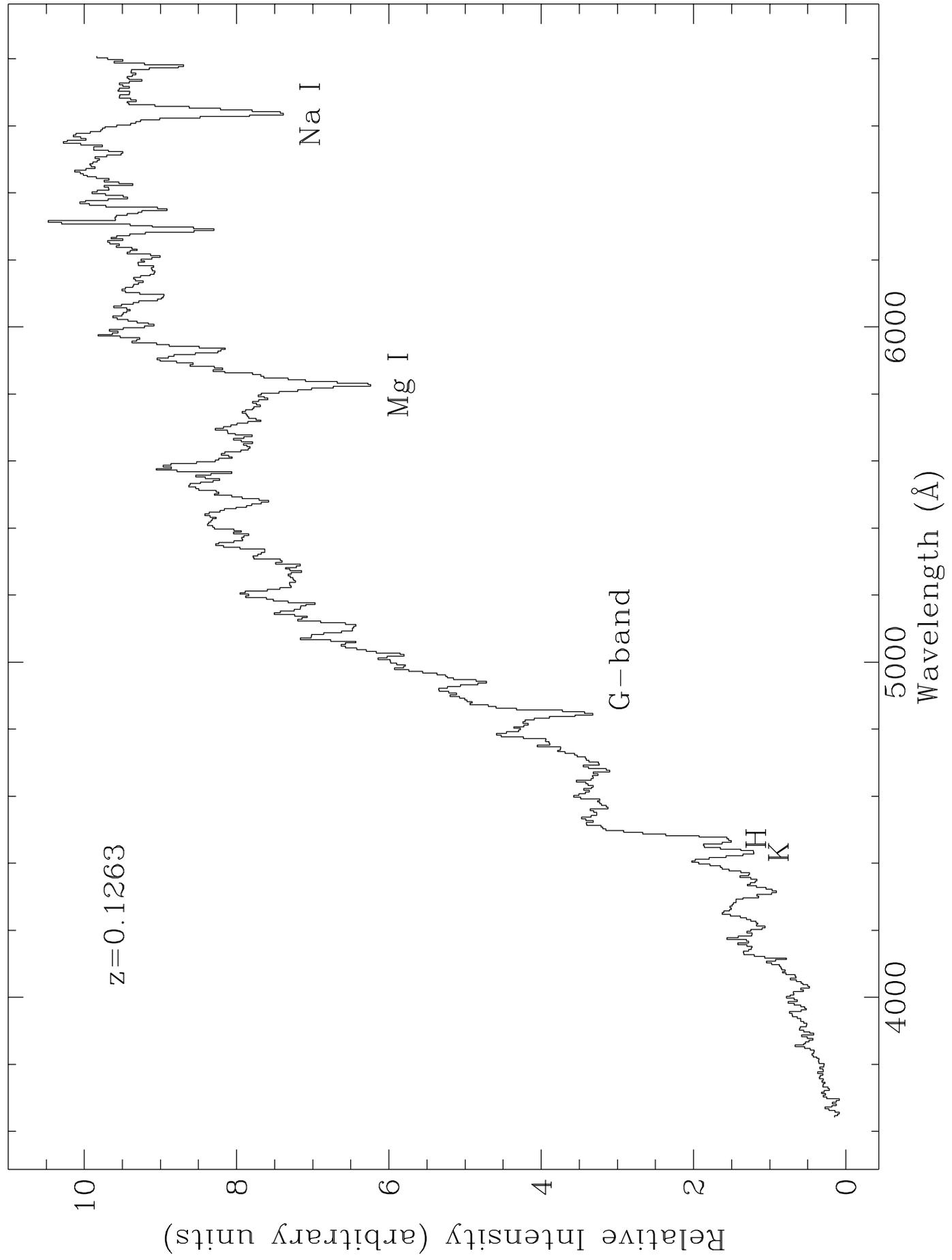

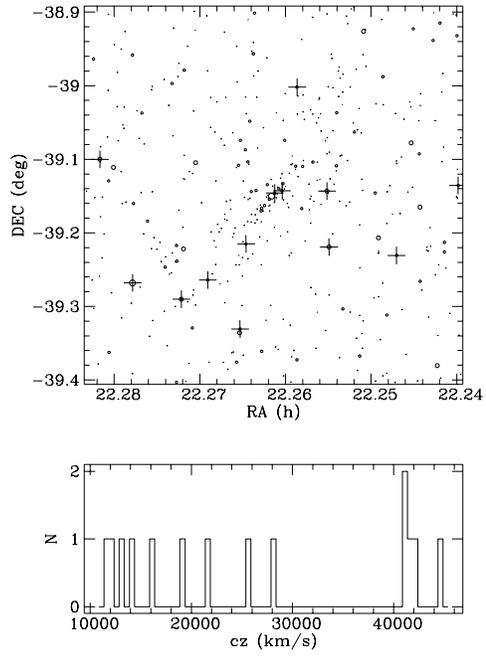

Figure 2. a

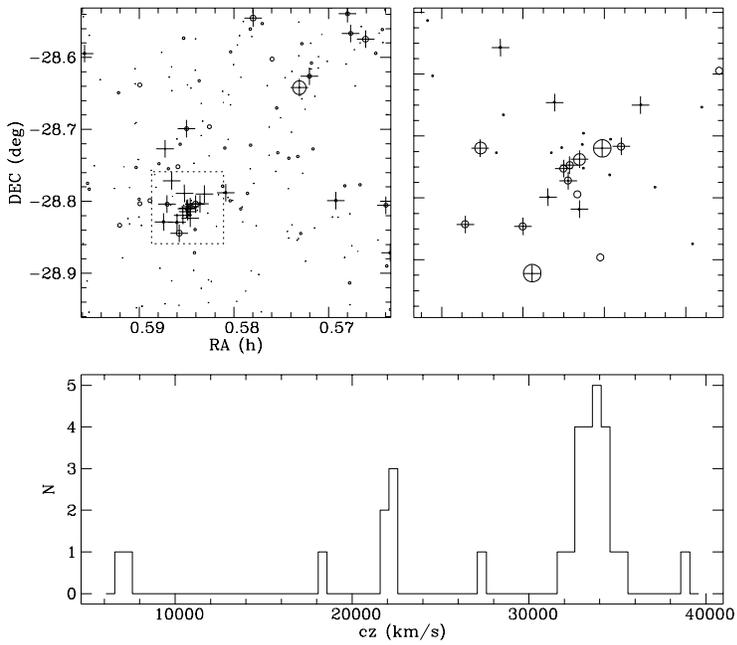

Figure 2. b

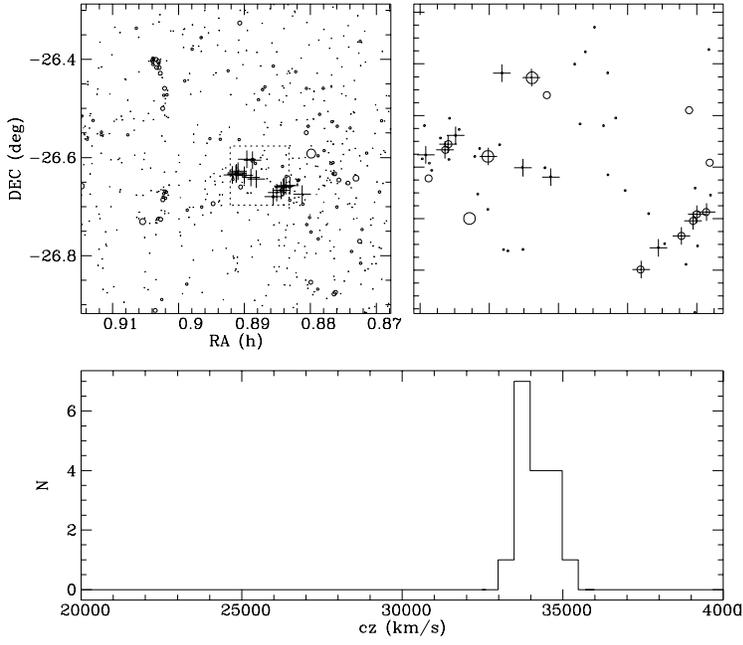

Figure 2. c

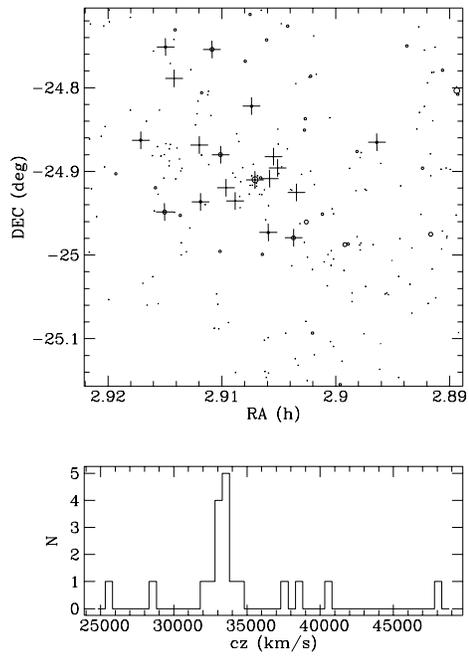

Figure 2. d

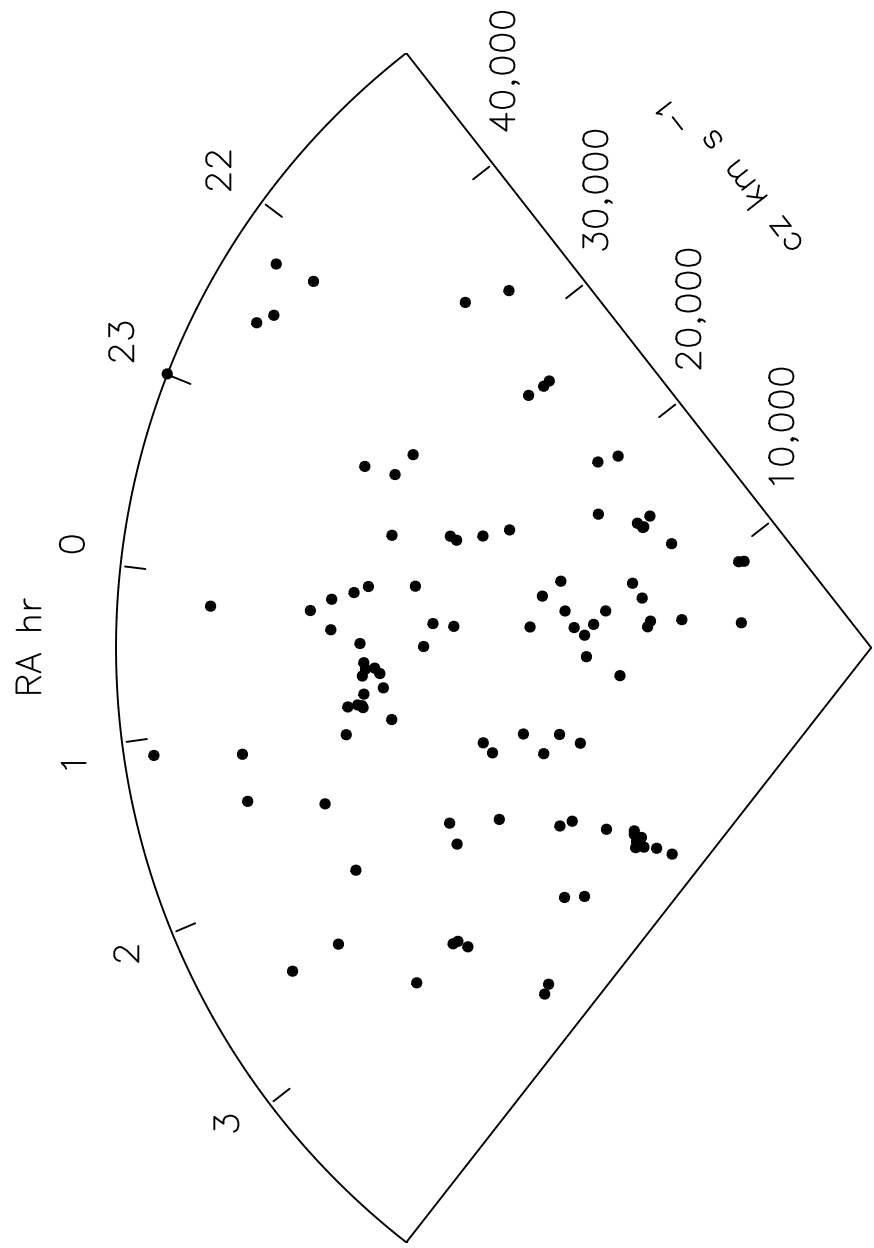

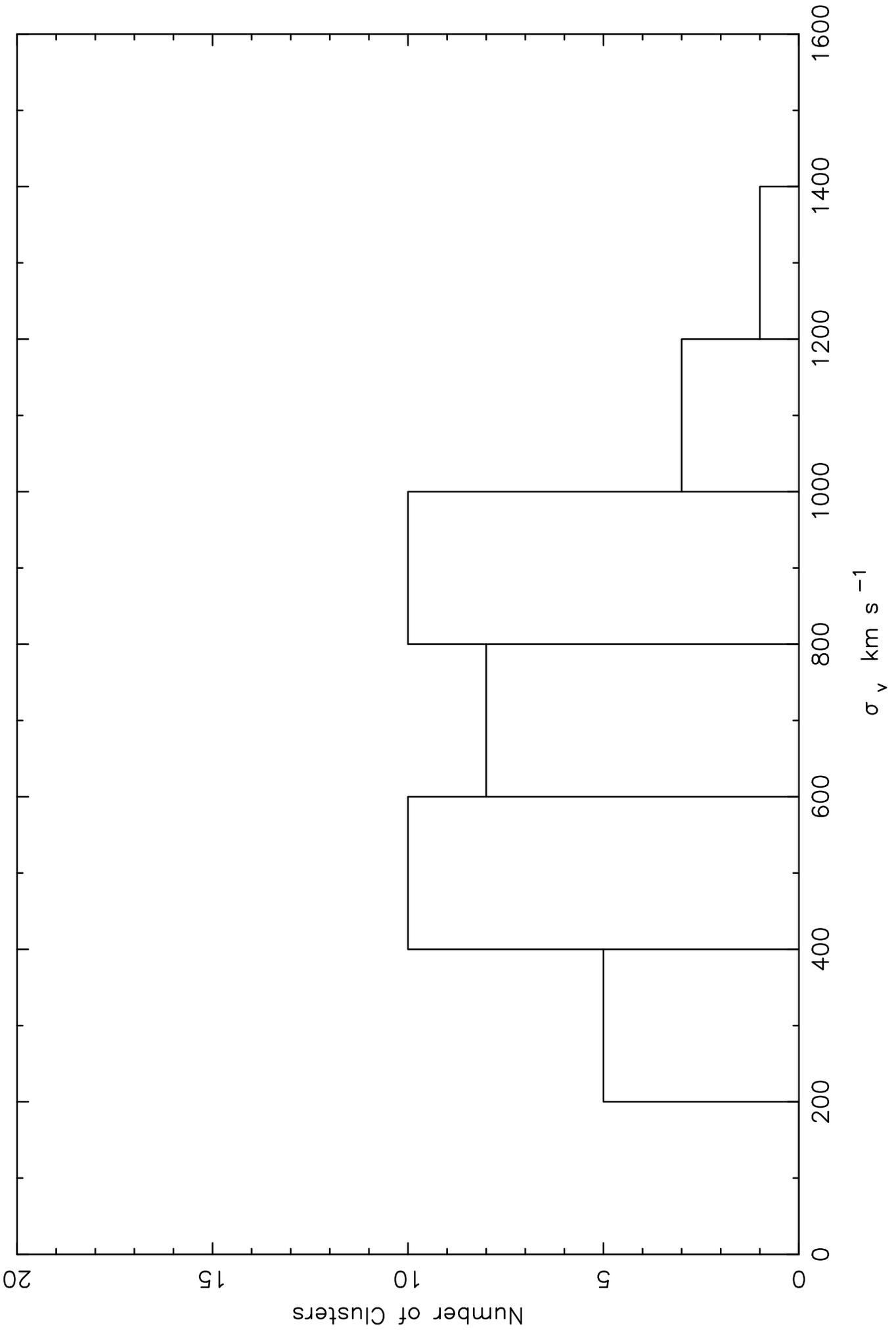